\documentclass[12pt,preprint,url]{aastex}
\usepackage{amsbsy}
\usepackage{amsmath}
\usepackage{color}
\newcommand{\be}{\begin{equation}}
\newcommand{\ee}{\end{equation}}

\def\ltsima{$\; \buildrel < \over \sim \;$}

\def\lsim{\lower.5ex\hbox{\ltsima}}
\def\gtsima{$\; \buildrel > \over \sim \;$}
\def\gsim{\lower.5ex\hbox{\gtsima}}

\shorttitle{Nonparametric pulsar glitch statistics}
\shortauthors{Melatos et al.}

\begin{document}
\title{Nonparametric estimation of the size and waiting time
distributions of pulsar glitches}

\author{G. Howitt\altaffilmark{1}, 
 A. Melatos\altaffilmark{1}, 
 and A. Delaigle\altaffilmark{2},}

\email{ghowitt@student.unimelb.edu.au, amelatos@unimelb.edu.au}

\altaffiltext{1}{School of Physics, University of Melbourne,
Parkville, VIC 3010, Australia}

\altaffiltext{2}{Department of Mathematics and Statistics,
 University of Melbourne, Parkville, VIC 3010, Australia}
 
\begin{abstract}
\noindent 
Glitch size and waiting time probability density functions (PDFs)
are estimated for the five pulsars that have glitched 
most using the nonparametric kernel density estimator.
Two objects exhibit decreasing size and waiting time PDFs.
Their activity is Poisson-like, and their size statistics are 
approximately scale-invariant.
Three objects exhibit a statistically significant local maximum in the PDFs,
including one (PSR J1341$-$6220) which was
classified as Poisson-like in previous analyses.
Their activity is quasiperiodic,
although the dispersion in waiting times is relatively broad.
The classification is robust: it is preserved across a wide range of
bandwidth choices.
There is no compelling evidence for multimodality,
but this issue should be revisited  when more data become available.
The implications for superfluid vortex avalanche models of pulsar glitches are explored briefly.
\end{abstract}


\section{Introduction 
 \label{sec:non1}}
Rotational glitches are impulsive, irregularly spaced spin-up events observed in pulsars.
The discovery of glitches came soon after the birth of pulsar astronomy itself: the first glitch
was detected in the Vela pulsar in 1968
\citep{Radhakrishnan1969,Reichley1969}, and in 1969 the first glitch was discovered in the Crab pulsar
\citep{Nelson1970,Lyne1993}.
Glitches have been discovered through large-scale monitoring programs 
with multibeam receivers at the 
Parkes and Jodrell Bank Observatories 
\citep{esp11,yu13}. 
At the time of writing, 
504 (430) events have been detected in 187 (143) objects\footnote{
Up-to-date catalogues are kept by the Jodrell Bank Centre for Astrophysics at
{\tt http://www.jb.man.ac.uk/pulsar/glitches.html}
and the Australia Telescope National Facility (ATNF) at
{\tt http://www.atnf.csiro.au/research/pulsar/psrcat/glitchTbl.html}.
Numbers quoted in the text without (with) parentheses refer to the Jodrell Bank 
(ATNF) data.
Numbers are current as at 2018 May 28.
\label{foot:non1}
},
amounting to $\sim 10\%$ of the known pulsar population.
The glitch catalogues record the instantaneous fractional change in the pulse frequency 
at the time of the glitch, known as the glitch size, $\Delta \Omega / \Omega$, 
and the epoch when each glitch occurs.
From the epochs, we calculate the waiting time, $\Delta t$, as the difference between
the epochs of successive glitches.

Glitches are thought to be caused by sudden readjustments within neutron stars;
see \citet{has15} for a recent review of theoretical glitch models. 
Studying the statistical properties of glitches may therefore lead to new insights into the physics of dense nuclear matter.
Previous statistical studies can be split into two categories: those that examine the population of glitching 
pulsars in aggregate, and those that examine the properties of individual pulsars. 
In the first category, we mention the work of \citet{mor93a}, who used size data to fit a power law
to the energy released during glitches,
and \citet{lyn00}, who looked at a sample of 32 glitches in 15 pulsars reported in \citet{she96} and found a correlation
between glitch activity and spin-down rate.
More recently, \citet{fue17} studied a sample of 384 glitches in 141 pulsars and found a correlation between 
glitch activity and the spin-down luminosity of the pulsar.
\citet{fue17} also examined glitch sizes in aggregate and found evidence for a multimodal
size probability density function (PDF) by fitting a mixed Gaussian model to the histogram of the data.
A similar analysis (with similar results) was performed by \citet{kon14}.
In the second category, we mention the work of \citet{mel08}, who constructed the empirical
cumulative density functions (CDFs) of glitch sizes and waiting times for the nine pulsars with the 
most recorded glitches and tested for consistency with avalanche models of glitch activity
\citep{war11,war12,war13}.
\citet{mel08} found that two pulsars, PSR J0537$-$6910 and PSR J0835$-$4510, 
differ from the rest of the pulsars studied in both their size 
and waiting time PDFs, a finding also reported in \citet{esp11}.
\citet{onu16} performed a similar analysis on `microglitches' (jumps in pulsar frequency with
$\vert \Delta \Omega / \Omega \vert \sim 10^{-10}$) in 20 pulsars using data from the Hartebeesthoek
radio telescope and again interpreted the results in the context of avalanche processes.
Studies of the most active glitching pulsar, PSR J0537$-$6910,
reveal a strong linear correlation between size and waiting time to the following glitch
\citep{mid06,fer17,ant18}.
Similarly, \citet{sha18} claimed to find a correlation in PSRJ 0534+2200 between size and waiting time since the previous
glitch.
\citet{eya17} found that the size PDF in 12 pulsars is well fitted by a normal distribution.

Statistical glitch studies usually posit
functional forms for the size and waiting time distributions,
e.g. finite mixture models
\citep{kon14},
or normal distributions
\citep{eya17}.
These methods assume that a set of global parameters define the PDF across its domain.
By contrast, a nonparametric estimator makes no assumptions about the global form
of the PDF but instead estimates the local probability density
around each data point
\citep{hal92a}\footnote{
The astrophysicist reader may recognize this as the fundamental idea behind 
smoothed particle hydrodynamics
\citep{mon05}.
\label{foot:non2}
}.
Nonparametric estimators are widely used in a host of scientific applications
\citep{sil86}.
Recently, nonparametric estimation has been used as an independent way to verify the discovery of two 
interesting new features in the Crab pulsar: a resolved minimum glitch size
\citep{esp14},
and an 11-year episode of accelerated glitch activity 
\citep{lyn15}.
In another recent application of nonparametric estimation in pulsar glitch research, 
\citet{ash17}
estimated the size distribution of all glitches to assess 
whether glitches will impact the discovery of gravitational waves from
rotating neutron stars.
In this paper, we study the individual glitch size and waiting time PDFs for the 
five pulsars with the highest number of recorded glitches using the kernel density estimator
\citep{wan95}.

We review the algorithm briefly, and evaluate its performance,
in \S\ref{sec:non2}.
We then apply the algorithm to construct waiting time and size PDF estimates 
for the most active glitchers 
in \S\ref{sec:non3a} and \S\ref{sec:non3b} respectively, and
the case for multimodality in certain objects is examined critically.
We compare the results with previous parametric studies and
astrophysical models in \S\ref{sec:non4}.
The results strengthen the empirical basis for theoretical work,
e.g.\ by firming up the identification of distinct classes of glitching pulsars.
They also offer a guide to designing the next generation
of glitch monitoring campaigns with phased radio arrays like LOFAR
\citep{kra10},
UTMOST 
\citep{cal16}
and 
the Square Kilometer Array.

\section{Kernel density estimator
 \label{sec:non2}}
 
\subsection{Definition
 \label{sec:non2a}}
 
We begin by defining the nonparametric kernel density estimator 
\citep{wan95}
we use to estimate the 
PDFs of glitch sizes and waiting times.
Let $x_1$, $\dots$, $x_N$ be $N$ independent, identically distributed samples
of a random variable $x$,
with underlying PDF $p(x)$.
The kernel density estimator ${\hat p}(x)$ of $p(x)$
is given by
\begin{equation}
 {\hat p}(x)
 =
 \frac{1}{N}
 \sum_{i=1}^N
 K\left(\frac{x - x_i}{h}\right)~,
\end{equation}
where $K(x)$ is a symmetric, positive definite kernel function $K(x)$
with normalization $\int dy\, K(y) = 1$.
The output of the kernel density estimator
is usually insensitive to the exact shape of $K(x)$.
Truncated polynomials and smooth functions multiplied by a Gaussian 
are common choices
\citep{wan95},
and these kernels produce an equivalent kernel density estimate under
an appropriate rescaling of the bandwidth
\citep{mar88}.
An exception is sharply peaked distributions, such as atomic spectra, where 
a different class of kernels known as `infinite order kernels' are more
appropriate
\citep{dav75,dav77}.
In this paper, we use a Gaussian kernel exclusively; 
the kernel corresponding to the datum $x_i$ is
\begin{equation}
K_i(x) = \frac{1}{h\sqrt{2 \pi}} \exp 
\left[ -\frac{1}{2} \left( \frac{x-x_i}{h} \right)^2 \right] \, .
\end{equation}
The value of the estimator at $x$ is a weighted tally of the observations 
$x_i$ in a neighborhood $| x - x_i | \lesssim h$ of $x$;
or, equivalently, it is the unweighted sum of $N$ identical copies
of the kernel function, centered at $x_1$, $\dots$, $x_N$.
Either way, each data point is spread across several bins
to give a smoothly differentiable PDF.
Kernel density estimation has achieved broad acceptance in many
applications because it represent a more optimal trade-off between bias and
variance than a bin-centred histogram. 
We also note that a bin-centred histogram is a kernel density estimator with a 
rectangular function as the kernel; in this work we use a Gaussian kernel so 
that the estimator inherits the useful properties of continuity and differentiability.

The key challenge when applying Equation (1)
is to select the bandwidth $h$ in a way that ensures good practical performance.
The ensemble-averaged, $x$-integrated bias
$\int dx\, \langle {\hat p}(x) - p(x) \rangle$
and variance
$\int dx\, \langle [ {\hat p}(x) - p(x) ]^2 \rangle$
increase and decrease respectively, as $h$ increases
in the limit $N\rightarrow \infty$.
Hence optimizing $h$ involves a compromise between bias and variance.
One approach is to let $h$ vary with $x$ according to
the local density of data points (see footnote \ref{foot:non2})
but it is ill-suited to small glitch samples.
An alternative, which has proven its worth in many applications,
is to choose a single, global $h$, that minimizes the asymptotic
mean integrated square error
\citep{wan95},
\begin{equation}
 {\rm AMISE } =
  \frac{1}{Nh} \int dx\, [ K(x) ]^2 
 +
  \frac{h^4}{4}
  \left[ \int dx\, x^2 K(x) \right]^2 
  \int dx\, [ p''(x) ]^2~.
\end{equation}
As $p''(x)$ is unknown a priori, it must be approximated.
Many techniques have been developed to estimate $p''(x)$ in order to select 
$h$ to minimise the AMISE [a summary of several of the more common techniques can 
found in \citet{wan95}]; 
in this work we use the normal reference bandwidth\footnote{This is, in effect, a second-order parametric assumption.
The normal reference bandwidth tends to produce an `oversmoothed' estimate,
sacrificing bias in order to reduce variance, and is more appropriate in the small-$N$
regime than other techniques such as Sheather-Jones plug-in
\citep{she91}
and smoothed cross-validation
\citep{hal92b}.}, which
assumes that $p(x)$ is a Gaussian with variance $\sigma^{2}$. 
In this case, the $h$ that minimises the AMISE can be written analytically, 
\begin{equation}
h =
\left\lbrace\frac{8 \pi^{1/2} \int dx \, [K(x)^2]}
{3N [\int dx \, x^2 K(x)^2]} \right\rbrace^{1/5} \sigma\, .
\end{equation}
The normal reference bandwidth rule replaces the true variance in equation (4) with 
some estimate $\hat{\sigma}^2$
\citep{sil86, sco92}. 
We take $\hat{\sigma}$ 
to be the minimum of the sample's interquartile range and standard deviation.
It is important to realise that the AMISE is, as the name implies,
an asymptotic measure.
An $h$ chosen to minimise the AMISE is just an approximation to the bandwidth that
truly optimizes the trade-off between bias and variance in the $N \leq 35$ 
regime of pulsar glitch statistics.
In \S\ref{sec:non2d} we explore the effect of bandwidth selection on synthetic data 
drawn from a bimodal distribution.
In \S\ref{sec:non3} we examine how the estimated PDFs change qualitatively with $h$.

\subsection{Positive definite variables
 \label{sec:non2b}}

If a PDF is defined on a finite domain and is non-zero at an endpoint, the kernel 
density estimator overspills the boundary.
Relative to an estimator that takes the boundary into account, the local bias 
$\langle \hat{p}(x) - p(x) \rangle$ 
in the vicinity of the boundary is greater than the $x$-integrated bias
$\int dx\, \langle {\hat p}(x) - p(x) \rangle$.
When Equation (1) is applied 
to a positive definite random variable,
such as glitch size or waiting time,
probability leaks spuriously into the region $x < 0$.
Leakage is significant, when a sizable fraction of the data points
satisfy $x_i \lesssim h$.
Renormalization does not fix the problem, 
as it introduces extra bias near $x=0$.
Many methods have been developed to counteract leakage.
Here we describe two procedures which are robust and straightforward to implement:
(i) 
reflect the data about $x=0$
and apply (1) with a symmetric kernel;\footnote{
One can also reflect about a right-hand boundary,
but this is irrelevant for glitches,
whose observed sizes and waiting times are much smaller than the maxima
that radio timing experiments can detect.}
or (ii) transform the data (e.g. by taking their logarithm).
Taking the logarithm compresses the domain of the size variable 
(which can otherwise extend over 4 dex in an individual object),
so that it can be modelled usefully by
a single, global $h$ chosen according to (2).
We apply the reflection method to the waiting time data in \S\ref{sec:non3a} and
the logarithmic transform method to the size data in \S\ref{sec:non3b}.

\subsection{Error estimation with nonparametric methods
 \label{sec:non2c}}

An obvious question when using the kernel density estimator is how accurate are the 
estimates produced? 
When using a parametric estimator, one can construct confidence intervals by varying
the parameters within a range and testing the likelihood of excluding the resulting
fit, e.g. with a K-S test as in Figures 2 and 3.
With the kernel density estimator, a similar approach is hindered by two issues:
the set of estimated parameters is large (formally the number of points where
the curve is estimated), and bias is one of the main contributors to the inaccuracy of 
the estimate (cf. parametric estimators, where the variance dominates)
\citep{hal92a,cha99}.
One approach to constructing confidence intervals that may seem appealing is 
to create data replicates by resampling and construct ${\hat p}$ for the
replicated data sets.
The simplest resampling `bootstrapping' method is inappropriate, because it produces a zero-bias 
confidence interval on the estimate ${\hat p}$ rather than on the true PDF
\citep{hal92a}.
Rather than seeking to construct a (potentially misleading) quantitative measure of the goodness of fit,
we instead take a qualitative approach, which aims to test whether particular features of the estimated
distribution, e.g. monotonicity and multimodality, are robust features that appear for a wide 
range of bandwidth choices
\citep{cha99,mar01}.
This concept is discussed further in \S\ref{sec:non2d} and \S\ref{sec:non3} below.

\subsection{Validation
 \label{sec:non2d}}
 
Before analyzing actual data, we run some tests on synthetic
data drawn from the exponential distribution $p(x) = e^{-x}$, which acts as a proxy 
for the glitch waiting time distribution, and the power law distribution
$p(x) = 0.2 x^{-1.2}, \, x \geq 1$
\citep{mel08}
The tests are not a substitute for 
definitive convergence studies presented elsewhere
\citep{sil86}
but give some sense of the reliability of the estimator and,
importantly, the sorts of artifacts (e.g.\ wiggles, plateaus) 
that arise from noise.
To construct kernel density estimates, we use the statistical software R
\citep{R},
and the \texttt{bkde} function included in the package \texttt{KernSmooth}
\citep{wan14}.

\begin{figure}
\begin{tabular}{c c}
\includegraphics[scale=0.35]{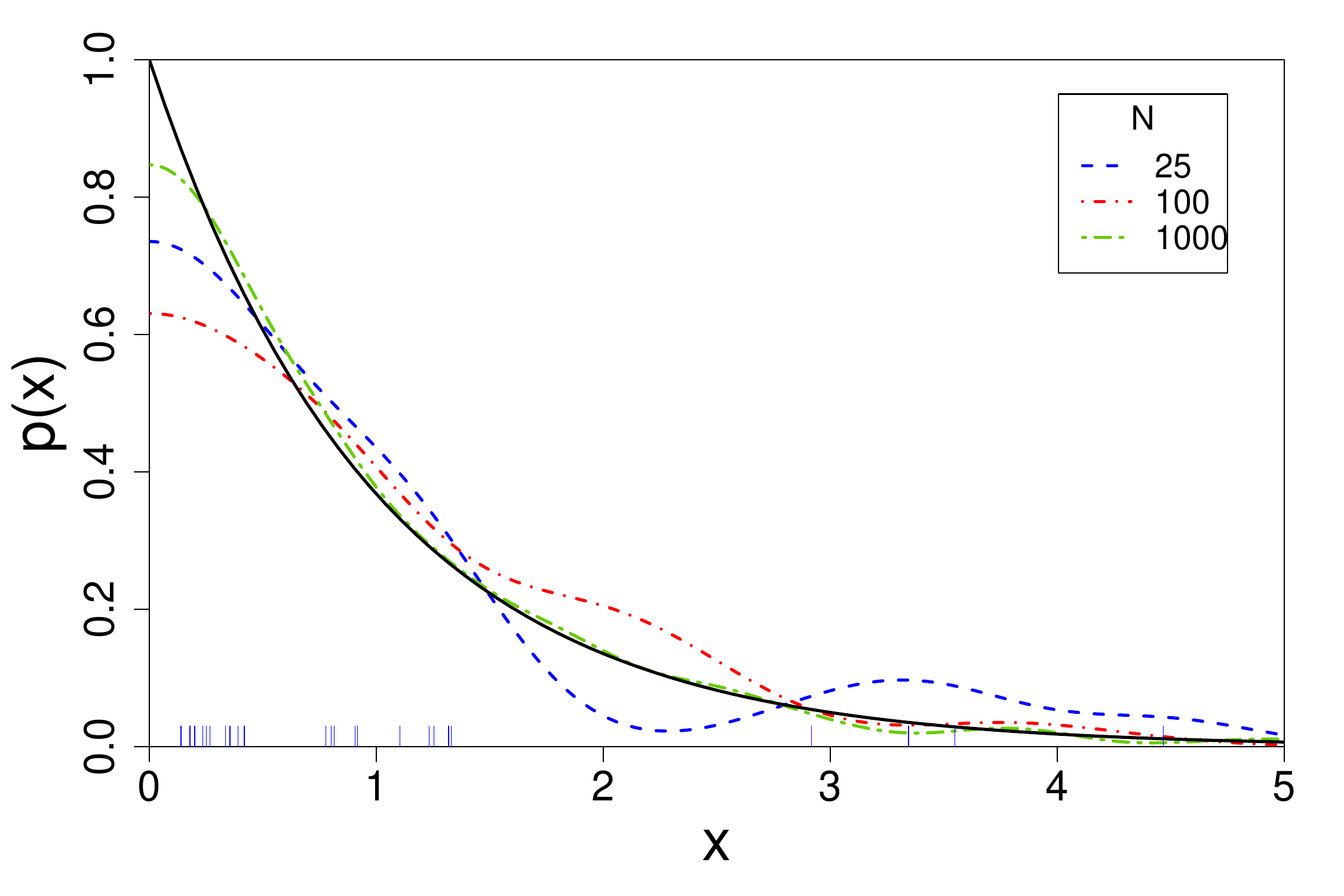} & 
\includegraphics[scale=0.35]{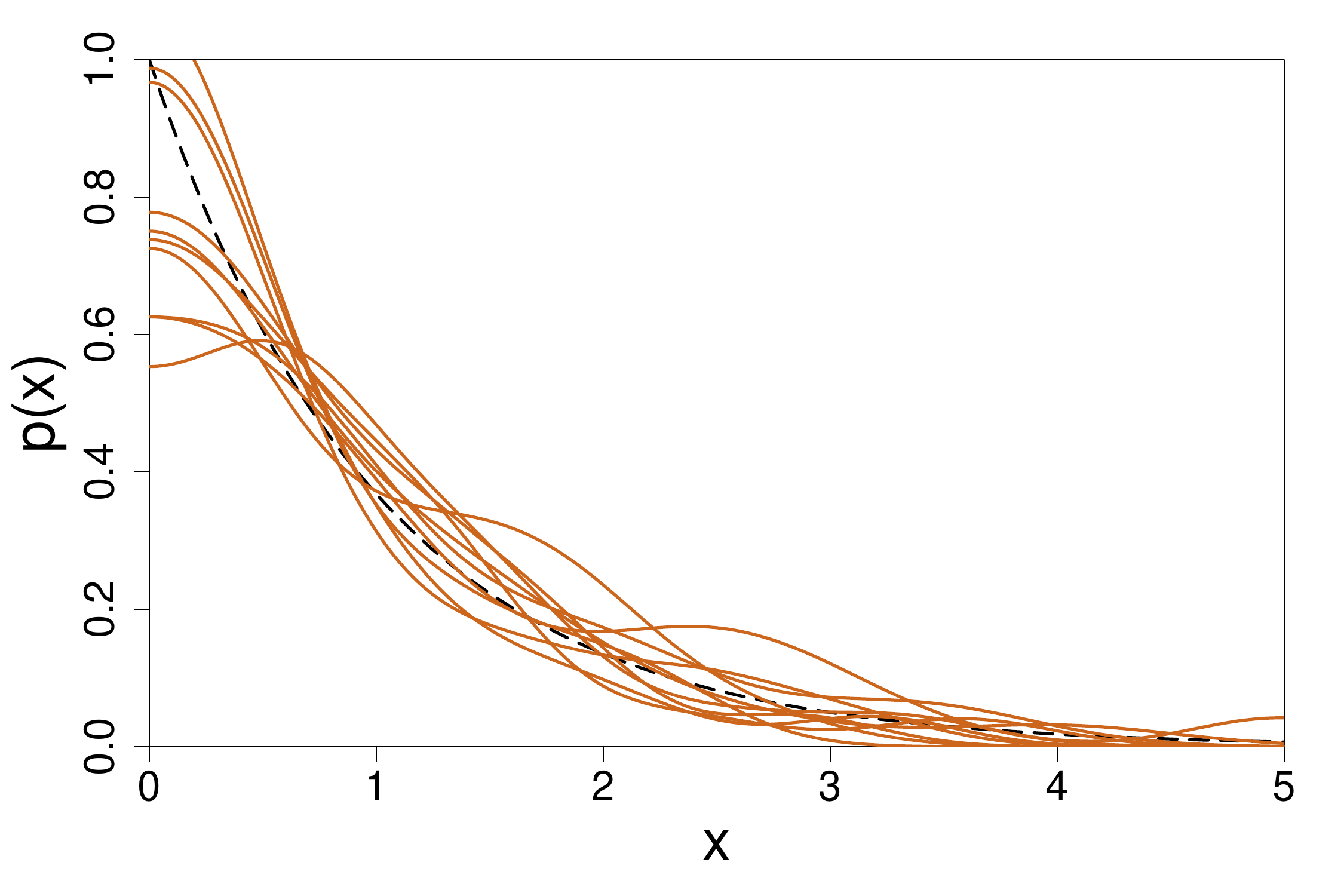} \\
\includegraphics[scale=0.35]{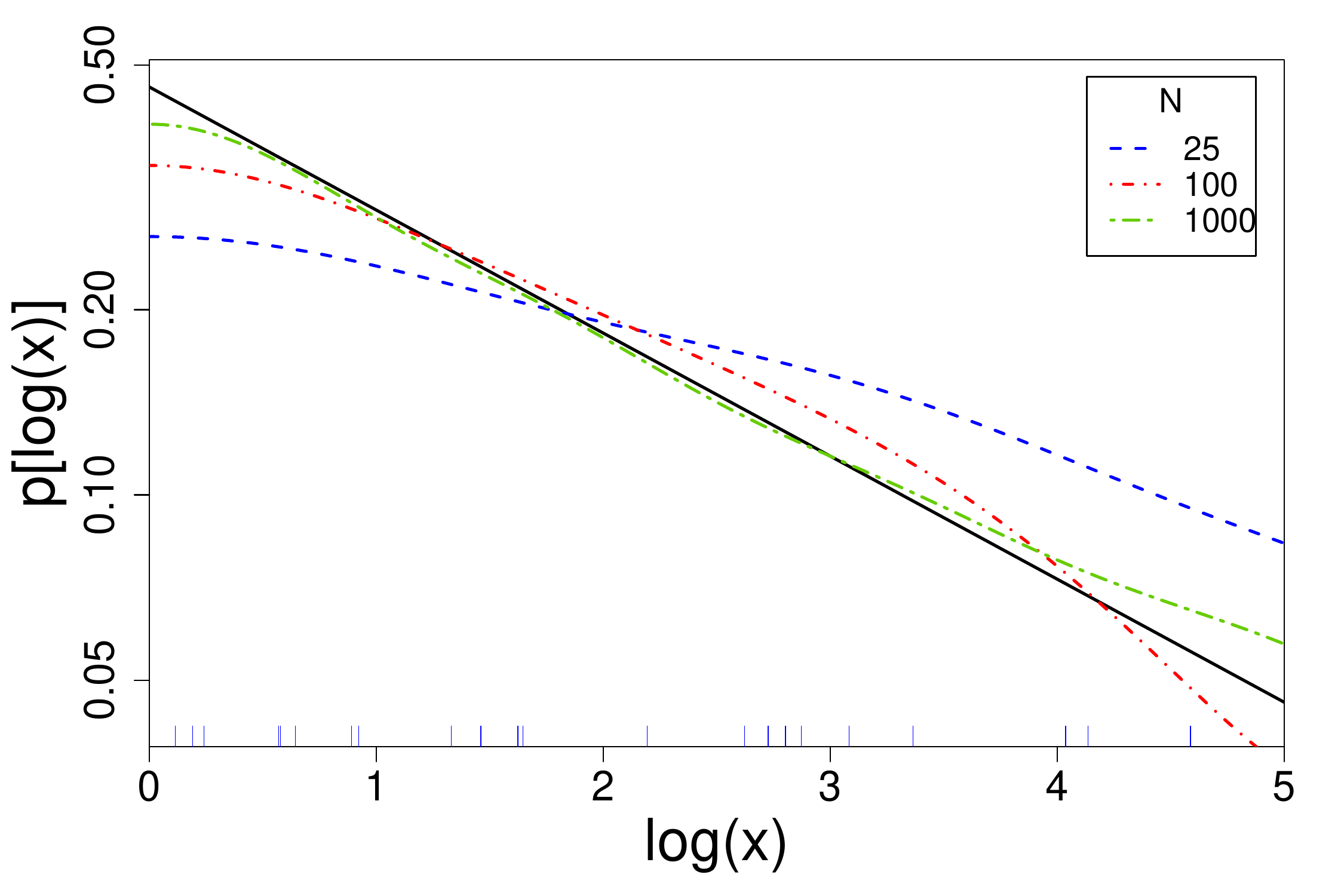} & 
\includegraphics[scale=0.35]{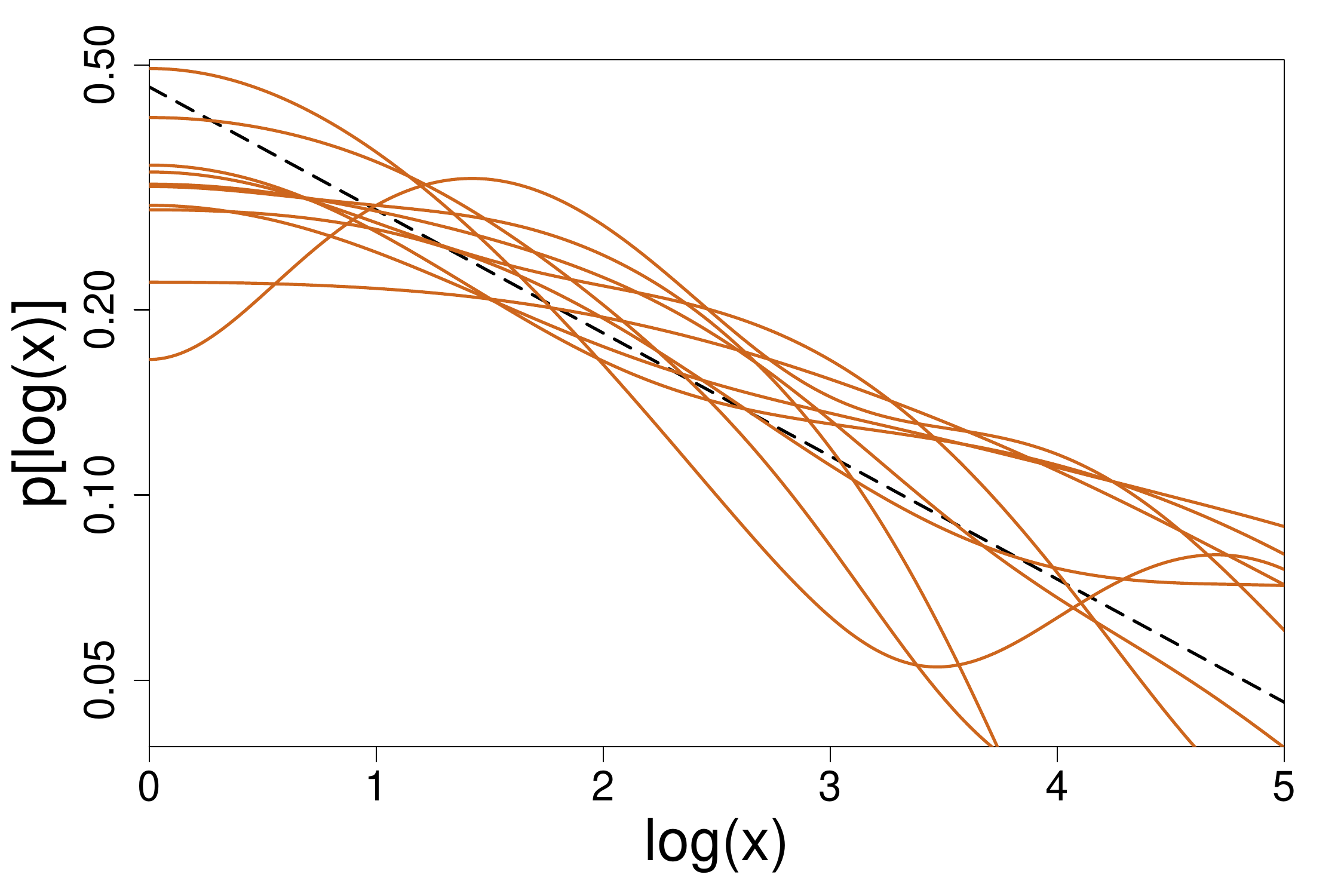}
\end{tabular}
\caption{
Convergence of the kernel density estimator ${\hat p}(x)$
[equation (1)] for an exponential PDF,
$p(x)=e^{-x}$ (top row),
and $\hat{p}[\log_{10}(x)]$ for the power law PDF,
$p(x)=0.2 x^{-1.2}$ with $x \geq 1$ (bottom row).
The solid black curve in the top row corresponds to the
exponential PDF,
The solid, colored curves in the left column
correspond to ${\hat p}(x)$ and ${\hat p}[\log_{10}(x)]$ for realizations with
$N=25$ (blue), $100$ (red), and $1000$ (green).
The tick marks in the panels in the left column indicate the abscissae
of the data in the $N=25$ (blue) realizations.
The dashed black line in the bottom row shows the power law PDF.
The right-hand column shows 10 instances of ${\hat p}(x)$ 
and ${\hat p}[\log_{10}(x)]$ with $N=25$ (gold curves). 
Equation (1) is applied to the reflected data in the top row
and the reflected base-10 logarithm of the data in the bottom row.
}
\label{fig:non1}
\end{figure}

Figure \ref{fig:non1} illustrates qualitatively 
how the estimator (1) performs with $N$,
when the underlying PDF is an exponential (top row) or a power law (bottom row).
For the exponential distribution, the reflection boundary correction 
described in \S\ref{sec:non2b} is used.
For the power law distribution, the kernel density estimate is applied to the
logarithm\footnote{If a variable $x$ is distributed according to a power law, i.e. 
$p(x) \propto x^{-a}$, 
then the PDF of the logarithmic variable
$y = \log_{10}(x)$ obeys $q(y) \propto (10^y)^{1-a}$.} of $x$.
Applying the estimate to $\log_{10}(x)$ reduces fluctuations 
(bumps and large gaps) that would otherwise appear, so that a 
global bandwidth gives good performance. 
As the domain is restricted to $x \geq 1$, and the probability density 
is significant at $x=1$, we apply the reflection method to the logarithmic data
at $\log_{10}x=0$.
In the top left panel, the colored curves show ${\hat p}(x)$
for three individual realizations 
with $N=25$, $100$, or $1000$ for the exponential distribution (solid black curve).
The bottom left panel shows $\hat{p}[\log_{10}(x)]$ for three realizations
with $N=25$, $100$, or $1000$, for the power law distribution (solid black line).
In the right column, each solid gold curve shows ${\hat p}(x)$ and 
${\hat p}[\log_{10}(x)]$ for one realisation with $N=25$, in order to give a 
sense of the scatter in the estimator.
We have checked many realizations and the results consistently exhibit
the following properties.
(i) Wiggles appear in ${\hat p}(x)$ for $N\leq 100$.
They are noise artifacts, which should not be interpreted as multimodality 
when analysing real data in \S\ref{sec:non3}.
(ii) Broadly speaking, the estimator performs similarly on the power law 
and the exponential.
(iii) The estimator plateaus at $x=0$, when the data are reflected (top row),
because $K$ being symmetric implies ${\hat p}'(0)=0$.
Hence ${\hat p}(x)$ systematically underestimates $p(x)$ near $x=0$,
if the underlying PDF is cuspy there,
although the bias decreases in a controlled fashion as $N$ increases\footnote{
In the absence of reflection, leakage can underestimate $p(x=0)$
by up to a factor of two, independent of $N$.
One can recognise this by noting that the positive and negative points 
in the reflected data contribute equally to $\hat{p}(x=0)$.
}.
In the top-right panel, $\langle \hat{p}(x = 0)/ p(x=0) \rangle = 0.71$, with a standard
deviation of $0.18$; in the bottom-right panel,
$\langle \hat{p}[\log_{10}(x)]/p[\log_{10}(x)] \rangle = 0.72$, with a standard deviation of 0.2.

\begin{figure}
\includegraphics[scale=1]{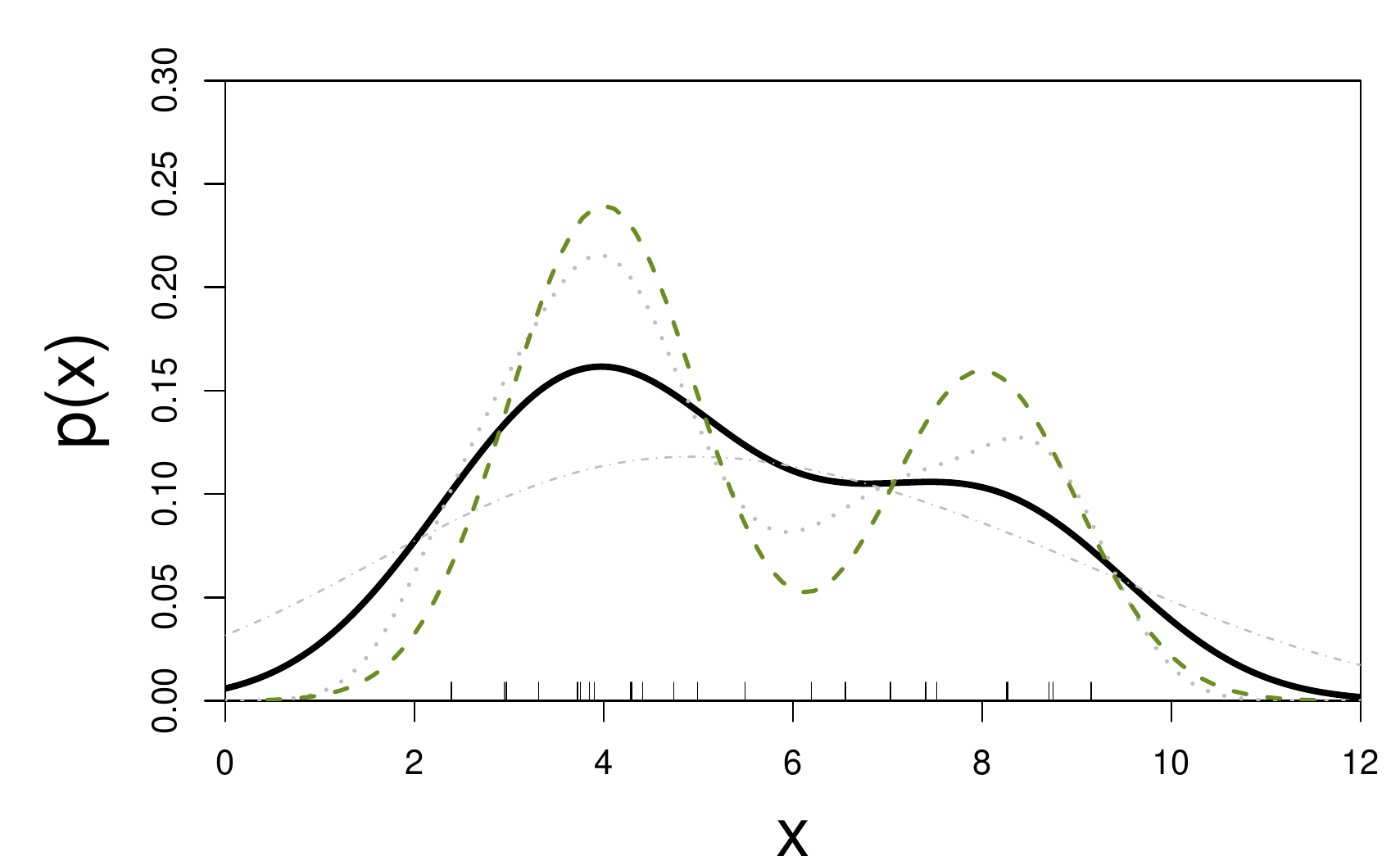}
\caption{Kernel density estimates of a two-component Gaussian PDF
 $p(x) = 0.6 \mathcal{N}(\mu = 4, \sigma=1) + 0.4\mathcal{N}(\mu = 8, \sigma = 1)$,
 with $N=25$ data points.
 The solid black curve shows the kernel density estimate with bandwidth $h_0$ chosen by the normal reference rule,
 the dotted grey curve is the kernel density estimate with bandwidth $h_0/2$ and the dot-dashed grey curve is with
 bandwdith $2h_0$. 
 The dashed green curve shows the underlying PDF from which the data used to construct the estimator,
 shown as black tick marks along the $x-$axis, are drawn. 
}
\label{fig:non2}
\end{figure}

Another issue worth addressing is the performance of the kernel density estimator for multimodal
PDFs, which have been proposed for glitch sizes in previous aggregated studies 
\citep{kon14,ash17,fue17}
and in analyses of individual pulsars 
\citep{mel08}.
We aim to determine whether the kernel density estimator can reliably reproduce multiple peaks in a PDF
as $h$ varies.
As a test, we consider a bimodal PDF obtained by summing two Gaussians, viz.
$p(x) = 0.6 \mathcal{N}(\mu = 4, \sigma=1) + 0.4\mathcal{N}(\mu = 8, \sigma = 1)$,
where $\mathcal{N}(\mu, \sigma)$ is the normal distribution with mean $\mu$ and standard
deviation $\sigma$.
We draw a variate of dimension $N = 25$ from $p(x)$, and construct the kernel density estimate with 
three choices of bandwidth: the normal reference bandwidth, $h_0$, as well as $h_0/2$, and $2h_0$.
The results are shown in Figure \ref{fig:non2}

Figure \ref{fig:non2} illustrates the characteristic property 
\citep{sil86}
of the kernel density estimator with the normal reference bandwidth: 
the features of the underlying PDF are recovered in the estimator,
which has local maxima at $x \approx 4$ and $x \approx 8$, but the estimator is somewhat `oversmoothed',
with $p(x)$ at the peaks underestimated by $\approx 30\%$ and $p(x)$ at the minimum overestimated by a factor of $\approx 2$.
The estimator with $h = h_0/2$ tends closer to the underlying PDF, however, the location of the peaks is less accurate.
The estimator with $h = 2h_0$ is excessively oversmoothed, as it completely fails to capture the bimodality in $p(x)$,
instead producing a single, broad peak centered about $x \approx 5$. 
As Figure 1 shows, it is possible for the kernel density estimator to produce local maxima even when the underlying
PDF is monotonic, so we tend to prefer the normal reference bandwidth with its tendency to slightly oversmooth. 
However, in our analysis of glitch data in \S\ref{sec:non3} we use the same set of bandwidths as in Figure 2 in order to illustrate 
the uncertainty inherent to the method
\citep{cha99}.

\section{Probability density functions
 \label{sec:non3}}
According to the Jodrell Bank catalogue at 2018 May 28, 504 glitches have been detected in 187 pulsars\footnote{
We work with the larger Jodrell Bank catalogue henceforth and cross-check against the ATNF catalogue.}.
\citet{yu17}
performed a detectability study of glitches in the 
\citet{yu13} data set and concluded that all detectable glitches in 
these data had been identified.
\citet{esp14} performed a detectability study of glitches in the Crab pulsar with Jodrell Bank data
and reached the same conclusion; see also \citet{jan06}.
Other than these two studies, however, no other authors have presented results on the completeness of glitch catalogues.
We stress that both the Jodrell Bank and the ATNF glitch catalogues may not be complete, and our analysis 
in this paper may be based on an incomplete data set.
In addition to the glitches in the Jodrell Bank catalogue, approximately 20 additional glitches have been reported in
PSR J0537$-$6910 in two papers by \citet{fer17,ant18}.
In this section, we construct size and waiting time PDFs 
for the five objects that have glitched most prolifically:
PSR J1740$-$3015 ($N=35$),
PSR J0534$+$2200 (26),
PSR J0537$-$6910 (42),\footnote{There is tension in the number of glitches and the parameters of glitches  for
PSR J0537$-$6910 between 
\citet{fer17,ant18} and earlier work by \citet{mid06}. 
We cross-check all three references and include the glitches common to two out of three sources in our analysis.} 
PSR J1341$-$6220 (23),
and
PSR J0835$-$4510 (21).
Experience across many scientific applications teaches 
that one needs $N\gtrsim 20$ to get meaningful nonparametric results
\citep{sil86},
as verified by Figure \ref{fig:non1}.
The sample sizes for the next most active objects, 
PSR J0631$+$1036 ($N=15$)
and
PSR J1801$-$2304 ($N=13$),
are too small to be analysed nonparametrically
[cf.\ \citet{mel08}];
their estimated PDFs are too bumpy.

\subsection{Waiting times
 \label{sec:non3a}}

\begin{figure}
\centering
\begin{tabular}{c c}
\includegraphics[scale=0.25]{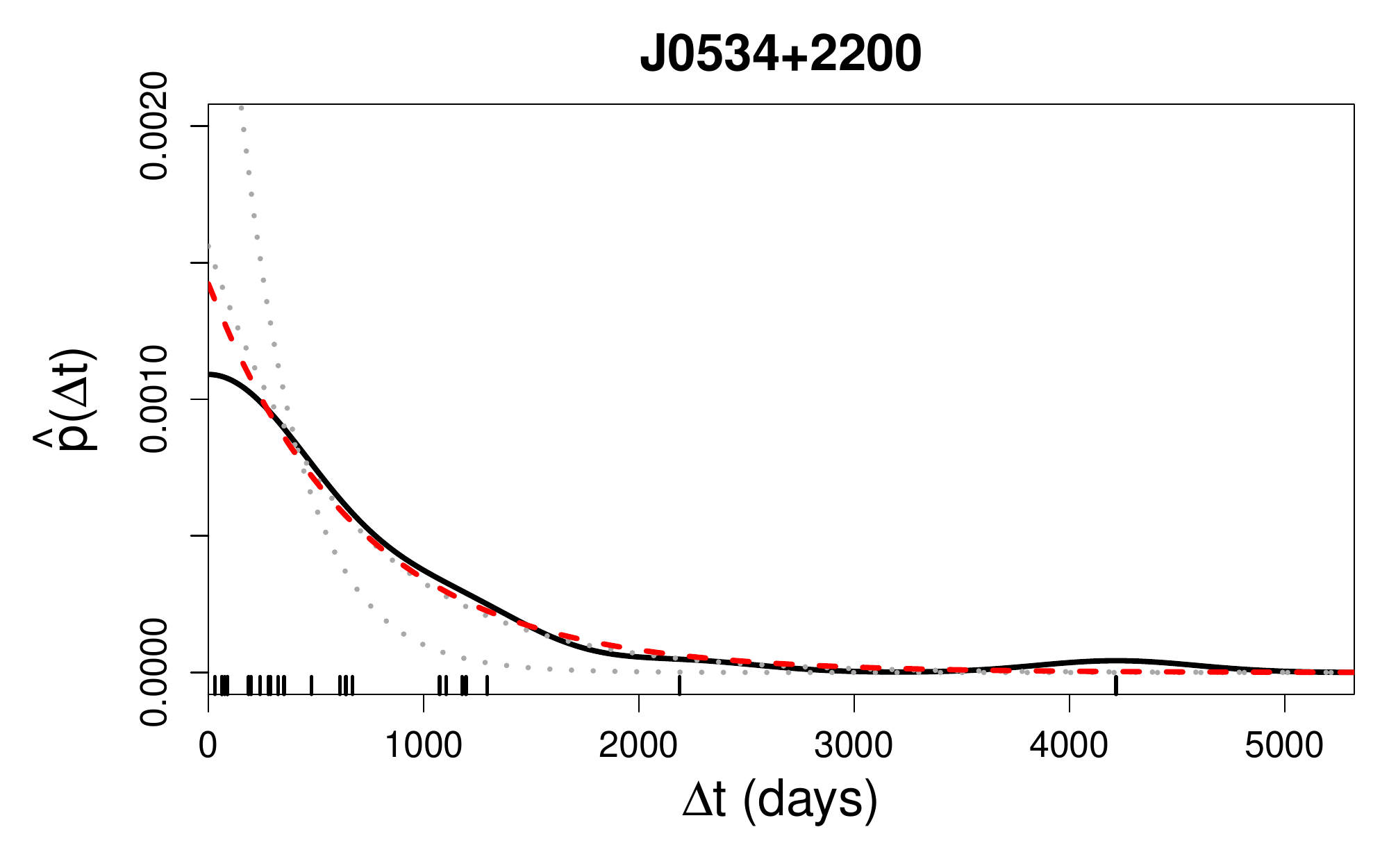} &
\includegraphics[scale=0.25]{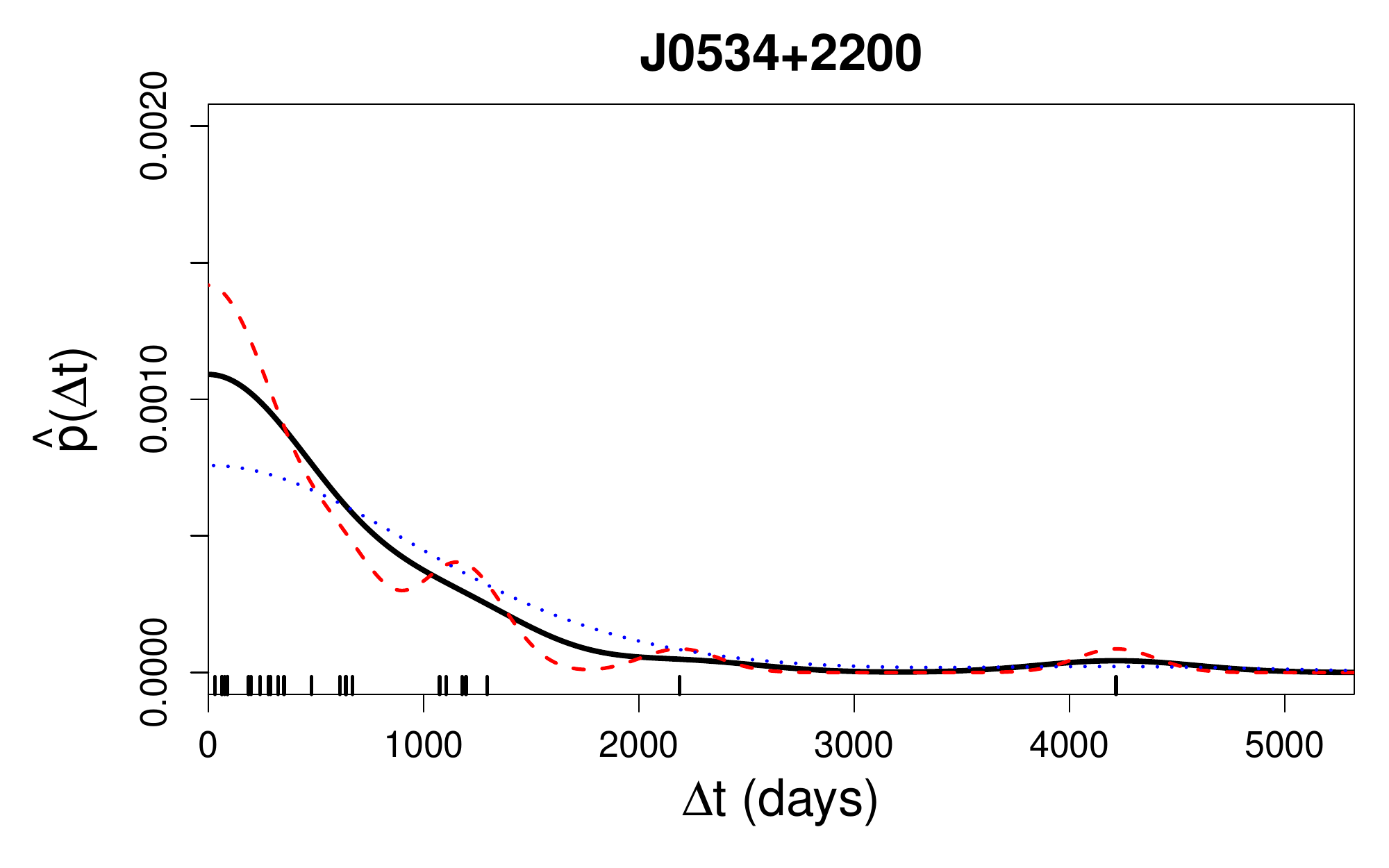} \\ 
\includegraphics[scale=0.25]{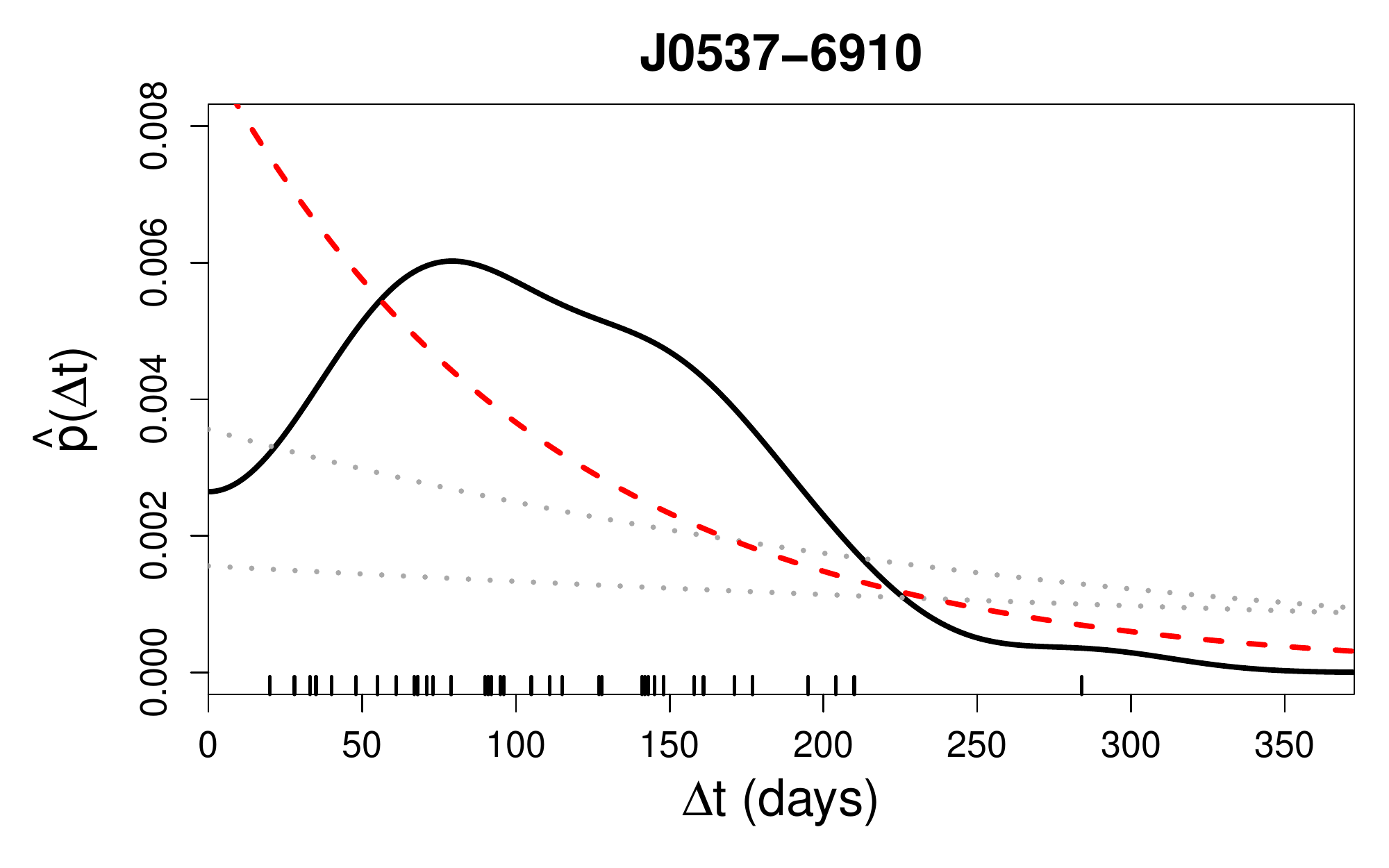} &
\includegraphics[scale=0.25]{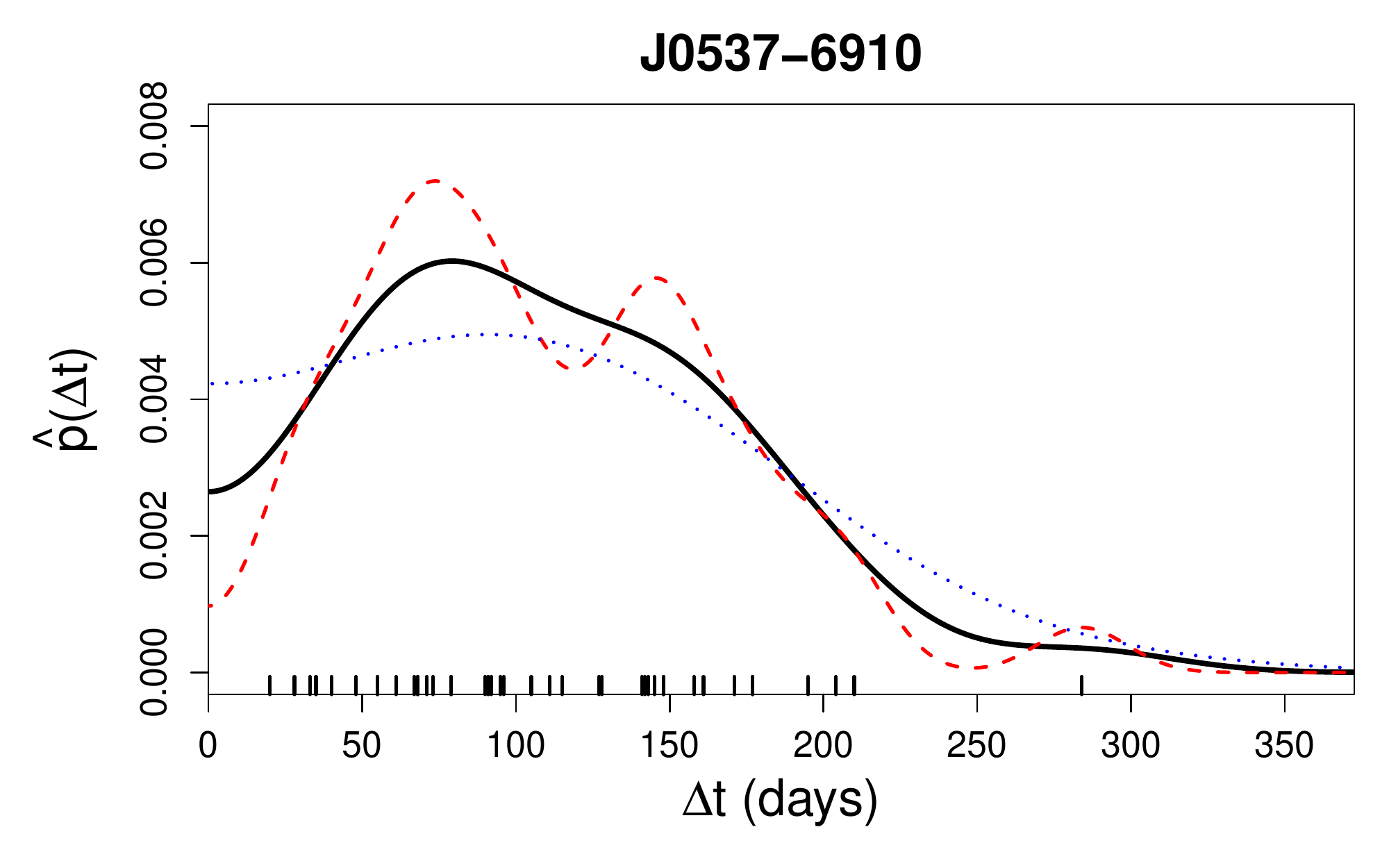} \\
\includegraphics[scale=0.25]{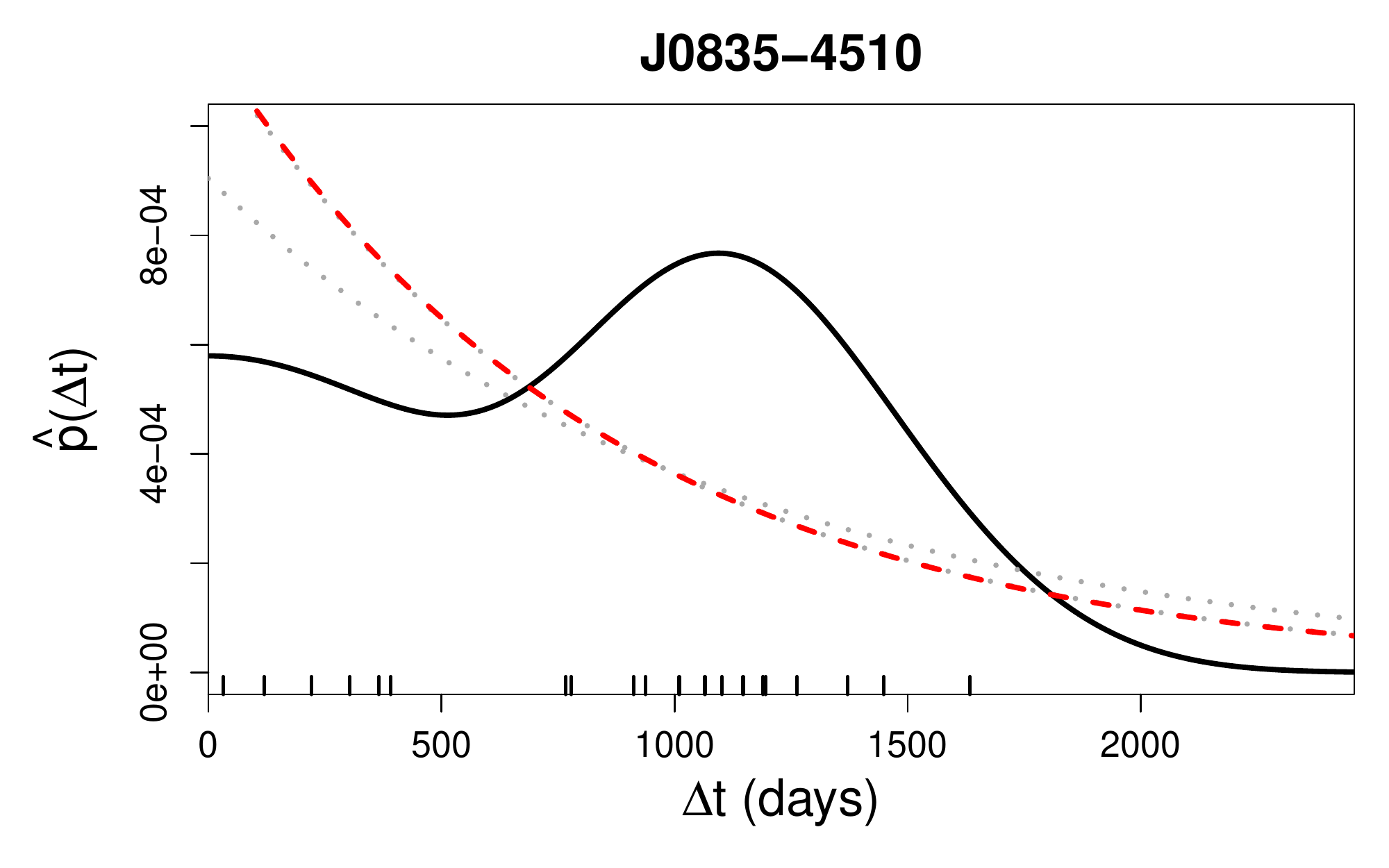} &
\includegraphics[scale=0.25]{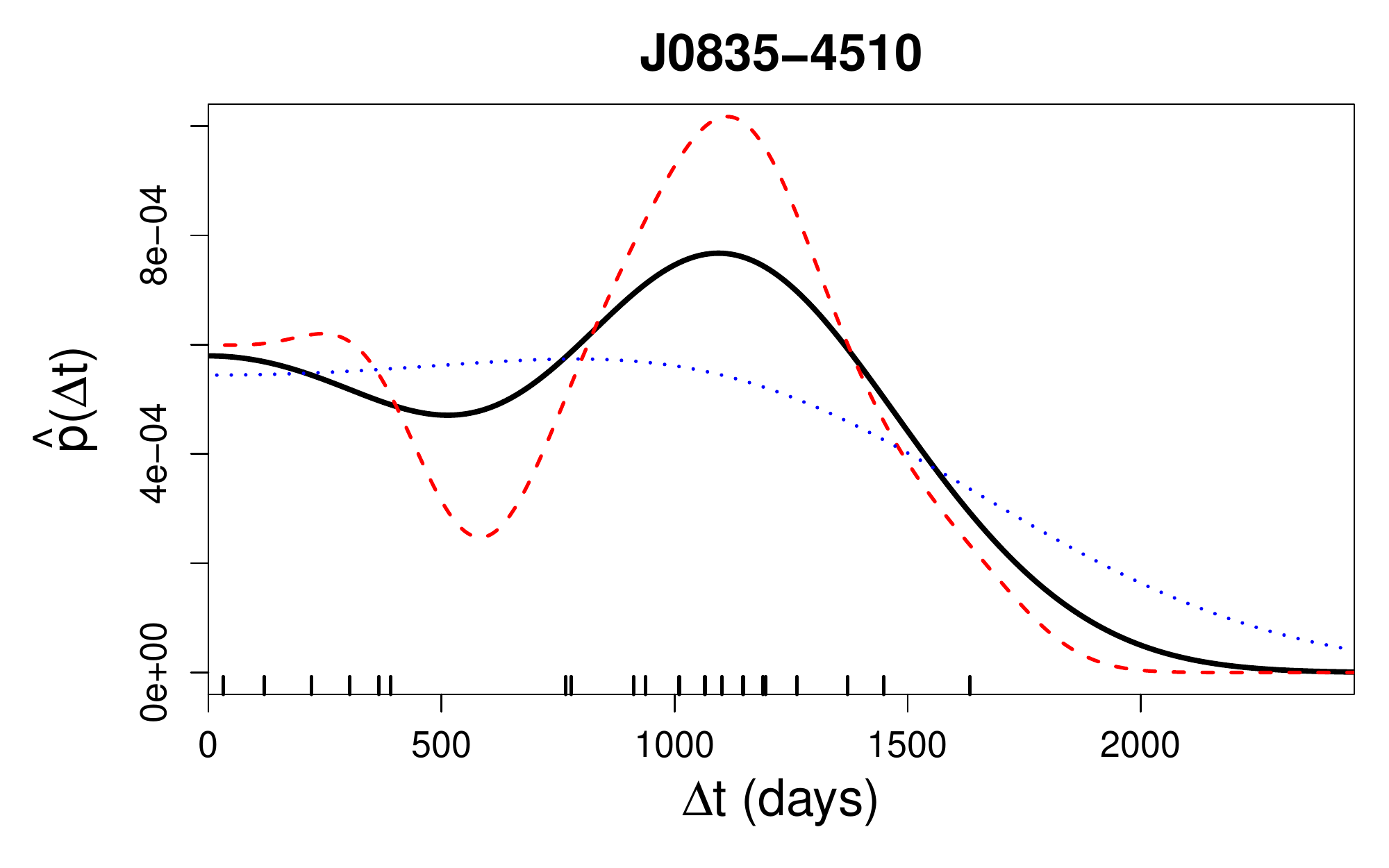} \\
\includegraphics[scale=0.25]{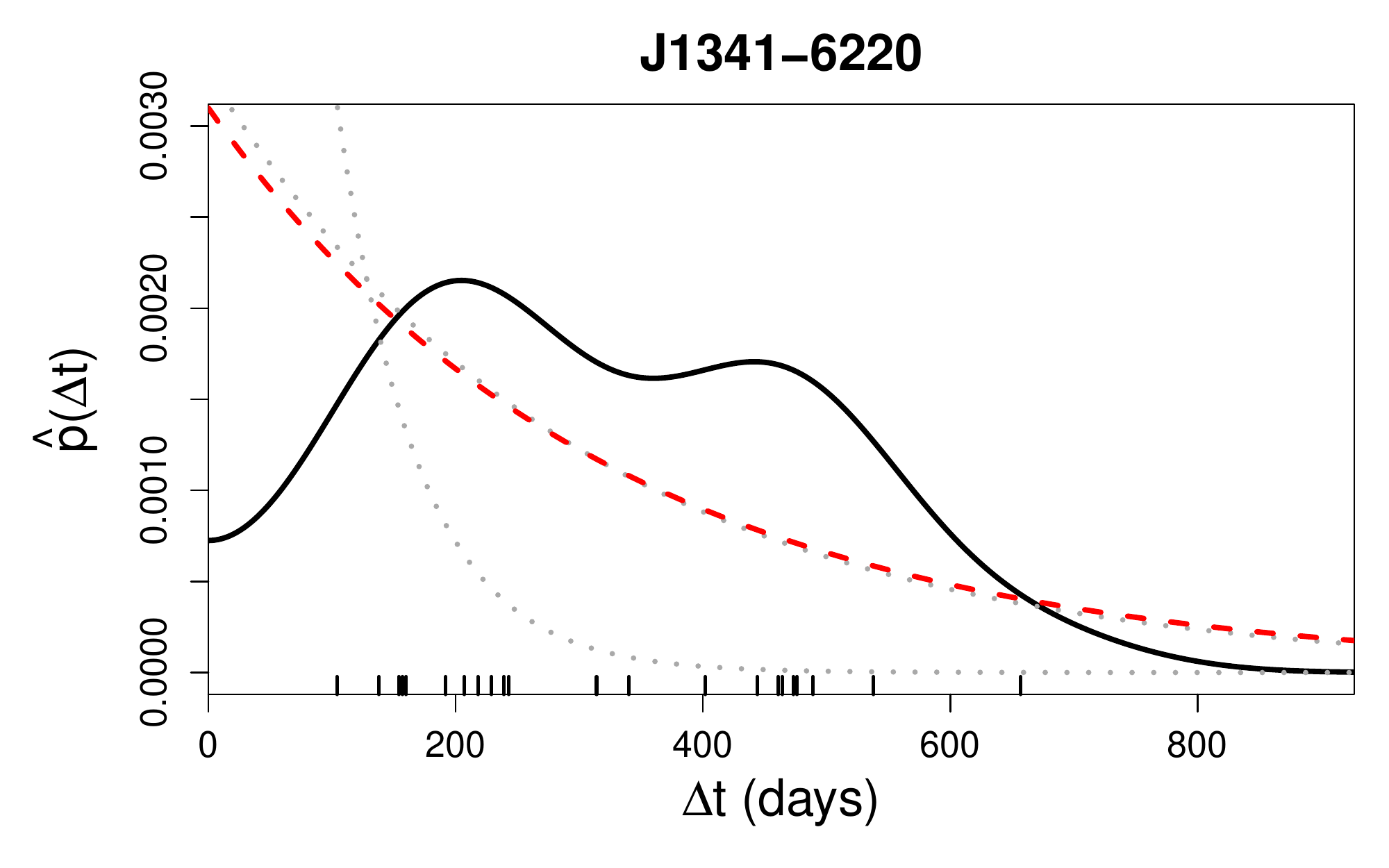} &
\includegraphics[scale=0.25]{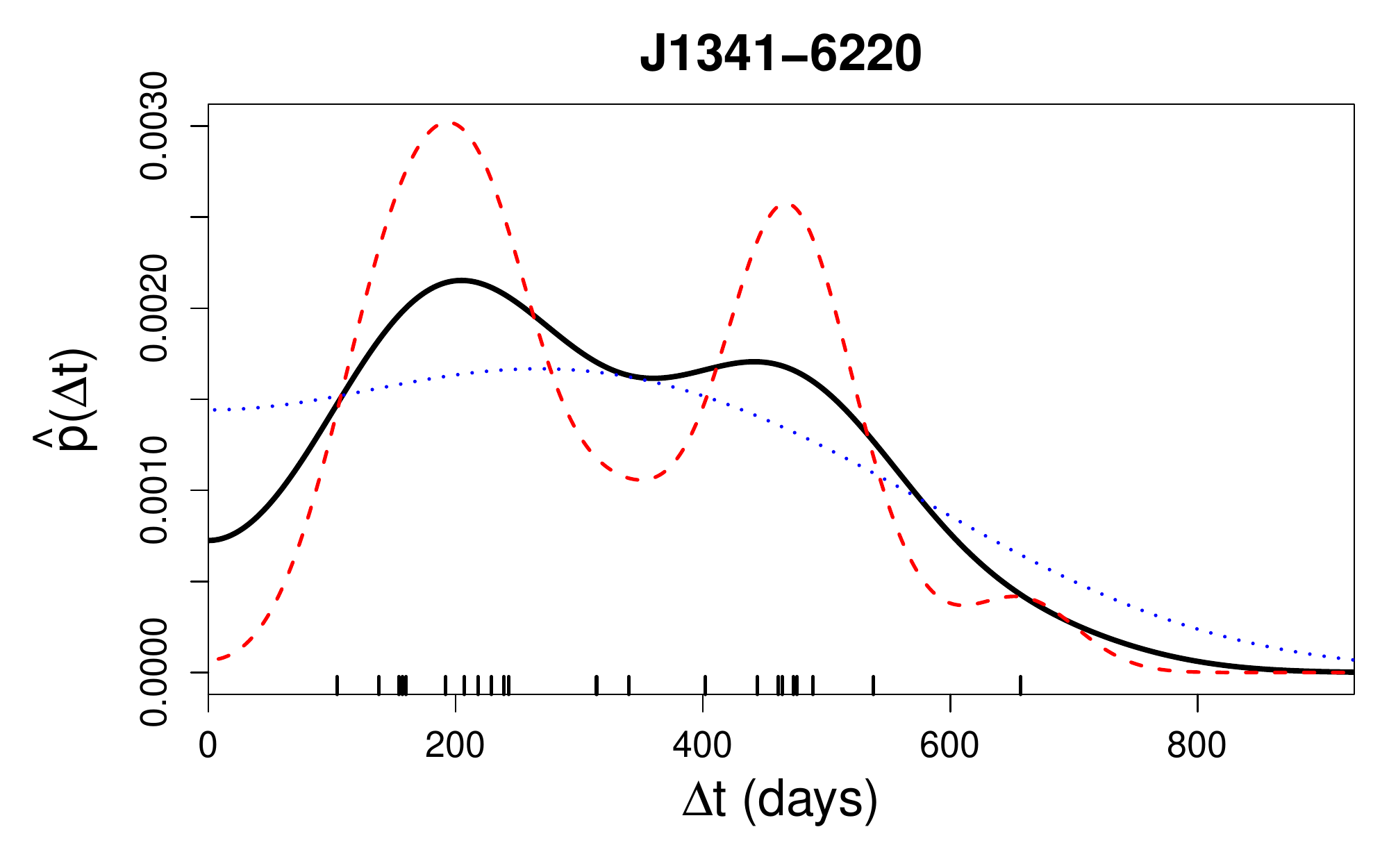} \\
\includegraphics[scale=0.25]{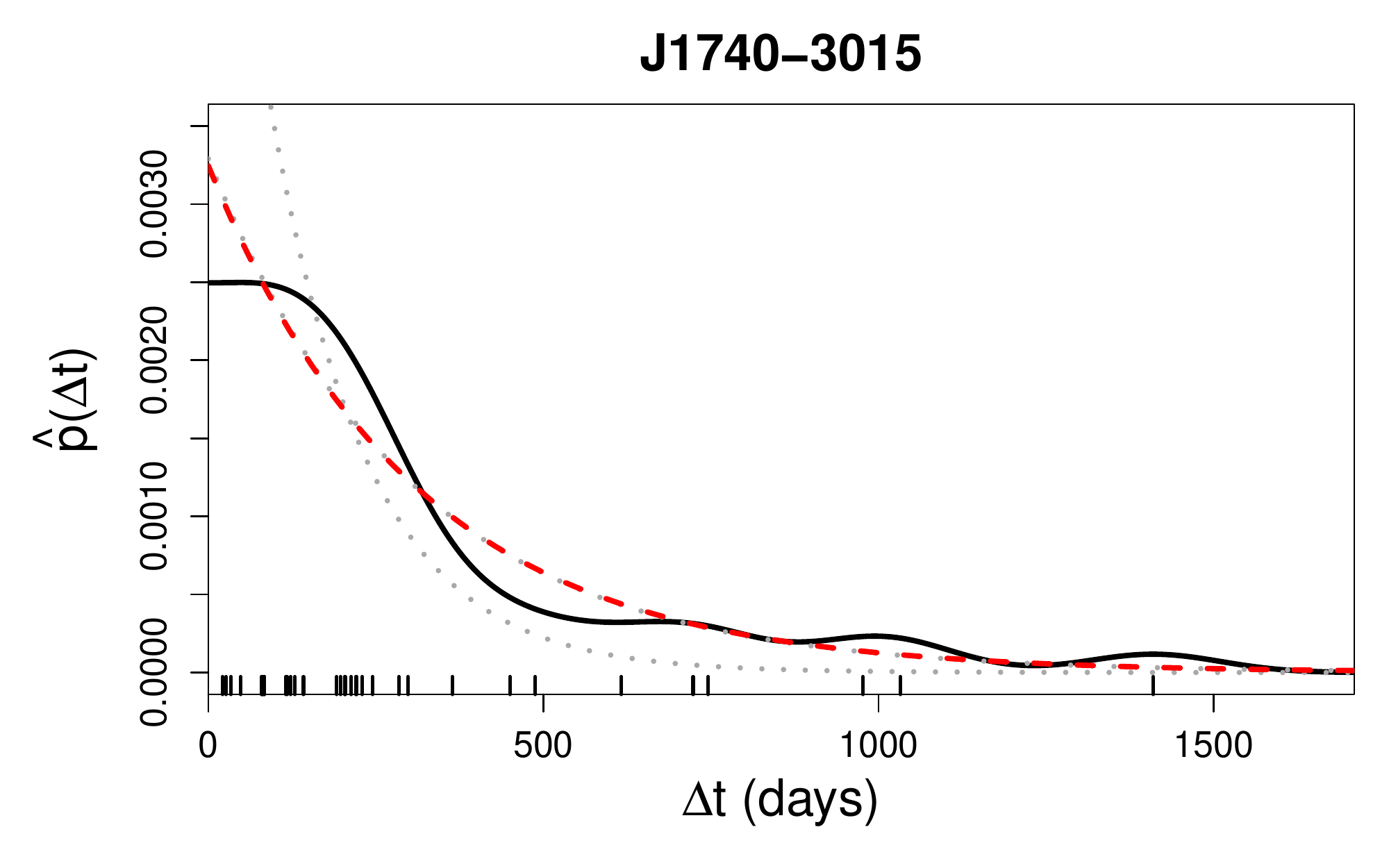} &
\includegraphics[scale=0.25]{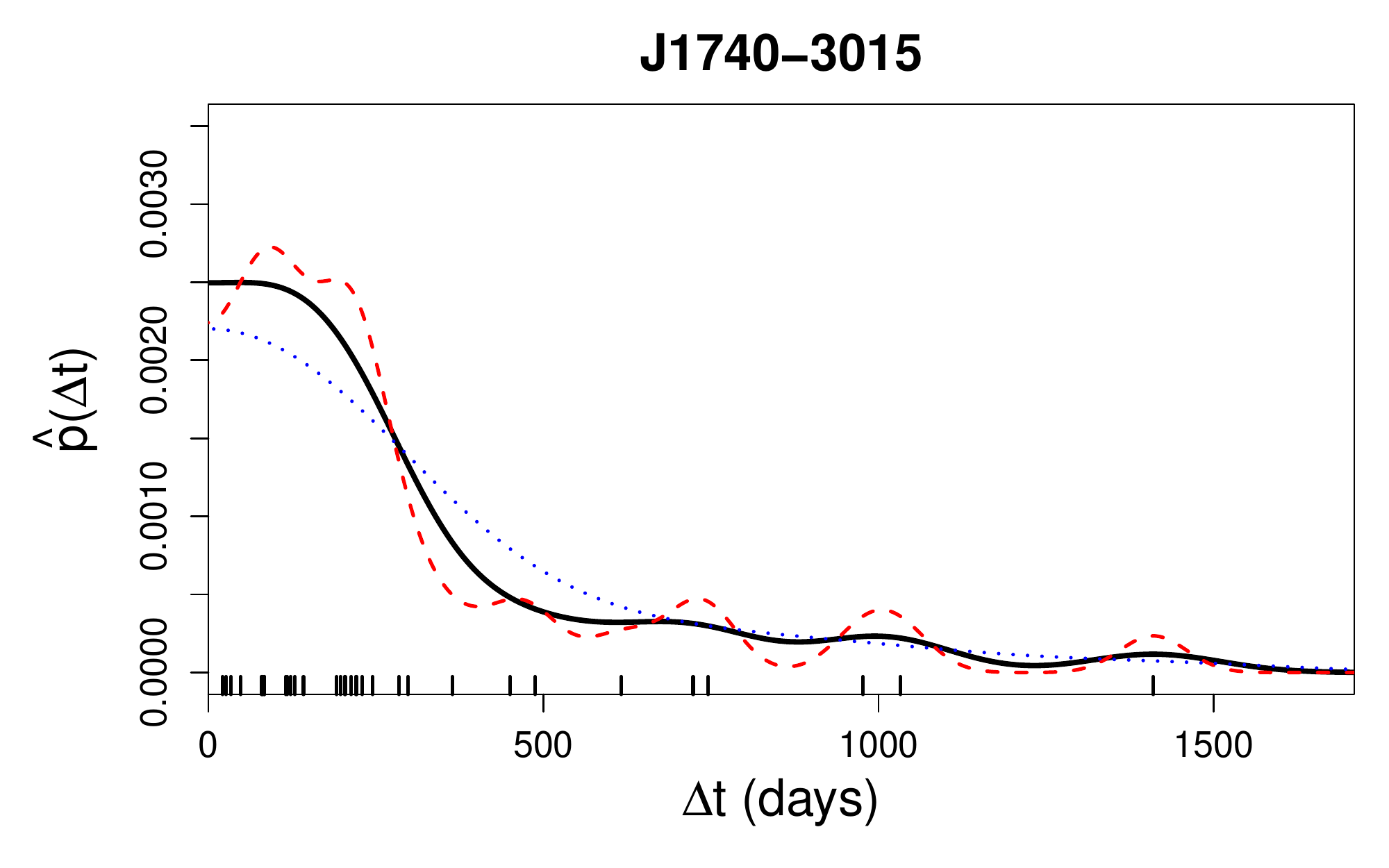} \\
\end{tabular}
\caption{
Kernel density estimates of the waiting time PDFs ${\hat p}(\Delta t)$ 
for the five most active glitchers, plotted on linear axes.
Solid curves correspond to the nonparametric kernel density estimator 
with a normal reference bandwidth $h = h_0$
In the left-hand column,
red dashed curves are parametric Poisson fits with 
$\lambda=\langle \Delta t \rangle^{-1}$ (maximum likelihood)
computed directly by averaging the data.
Grey dotted curves are Poisson fits with 
$\lambda=\lambda_\pm$ from the K-S analysis in 
\citet{mel08}.
In the right-hand column, the red dashed curves are the kernel density
estimator with bandwidth $h = h_0/2$, the blue dotted curve is with $h = 2h_0$.
Tick marks on the horizontal axis indicate the abscissae of the 
raw data.
}
\label{fig:non3}
\end{figure}

Figure \ref{fig:non3} displays the waiting time PDFs for the five pulsars 
selected above.
In each panel, the nonparametric kernel density estimator ${\hat p}(\Delta t)$
(units: yr$^{-1}$) from equation (1), using the normal reference bandwidth, $h = h_0$,
is graphed on a linear scale as a solid curve.
In the left column of \ref{fig:non3}, we displays fits to the Poisson PDF, 
$p(\Delta t) = \lambda e^{-\lambda \Delta t}$.
The red dashed curve corresponds to the maximum likelihood estimate 
$\lambda = \langle \Delta t \rangle^{-1}$.
The grey dotted curves correspond to 
the values of $\lambda$ which demarcate the interval 
$(\lambda_-, \lambda_+)$, where the null
hypothesis that the data are drawn from a Poisson distribution is rejected 
by a K-S test with less than $68\%$ confidence
\citep{mel08}.
In the right-hand column, we display two additional kernel density estimators using 
different bandwidths. 
The dashed red curve is with $h = h_0/2$, the dotted blue curve is with $h = 2h_0$.

Figure \ref{fig:non3} reveals three important things.
First, broadly speaking,
${\hat p}(\Delta t)$ decreases with $\Delta t$ for 
PSR J0534$+$2200 and PSR J1740$-$3015,
notwithstanding the bumps in the body and tail of the PDF estimate and the
low-$\Delta t$ plateau, 
which are both peculiarities of the kernel density estimator,
as implied by Figure \ref{fig:non1}.
It is apparent by eye that ${\hat p}(\Delta t)$ 
is broadly consistent with the parametric Poisson PDF
posited previously
\citep{won01,mel08,esp11};
the dashed curve in the left-hand column hugs the solid curve 
and falls between (and nearly coincides with) 
the dotted fits from \citet{mel08},
even after adding three events since 2008.

Second,
${\hat p}(\Delta t)$ rises to a maximum
in the other objects.
This is interesting, since PSR J0537$-$6910 and PSR J0835$-$4510 are traditionally classified 
as quasiperiodic glitchers \citep{mid06,mel08},
but PSR J1341$-$6220 was previously classified as Poisson-like 
\citep{mel08}.
The peak in ${\hat p}(\Delta t)$ in the latter object is broad;
in fact, the dashed Poisson curve is still a fair fit to the solid curve,
albeit displaced downwards.
However,
${\hat p}(0)$ is less than $30\%$ of ${\hat p}(\Delta t)$
at the peak for all three non-Poisson-like objects, and the maximum in $\hat{p}$
is at $\Delta t > 0$ using the most oversmoothing bandwidth $h = 2h_0$;
suggesting the existence of a maximum at $\Delta t > 0$ instead of $\Delta t = 0$

Third, there is some evidence of bimodality in PSR J0835$-$4510 and PSR J1341$-$6220, 
both of which show local minima in $\hat{p}(\Delta t)$ with $h = h_0$ and $h = h_0/2$.
Quasiperiodic glitchers have been modelled with a two-component
PDF in previous work
\citep{mel08,kon14,ash17,fue17}.
In particular, it is sometimes said that the `big' glitches in 
PSR J0835$-$4510 are spaced regularly, but the `small' ones
are not.
The short-$\Delta t$ component, if present, arguably
dominates the long-$\Delta t$ component in PSR J1341$-$6220,
whereas the situation is the other way around in PSR J0835$-$4510.
Looking at $\hat{p}(\Delta t)$ for PSR J0534+2200 and PSR J1740$-$3015, 
with $h = h_0/2$, however, shows that this choice of bandwidth significantly
`undersmooths' the PDF, producing multiple peaks centred about one or two
events. 
On balance, we do not see convincing evidence for bimodality
in ${\hat p} (\Delta t)$ in Figure \ref{fig:non3}.
The gentle bumps in Figure \ref{fig:non3} look much like
the small-$N$ features identified in the validation experiments 
in Figure \ref{fig:non1} due to sparse sampling of the PDF.
More data are required to rule out bimodality definitively,
but there are certainly no strong grounds for claiming its
existence on the basis of a nonparametric analysis at present.

\subsection{Sizes
 \label{sec:non3b}}
\begin{figure}
\centering
\begin{tabular}{c c}
\includegraphics[scale=0.25]{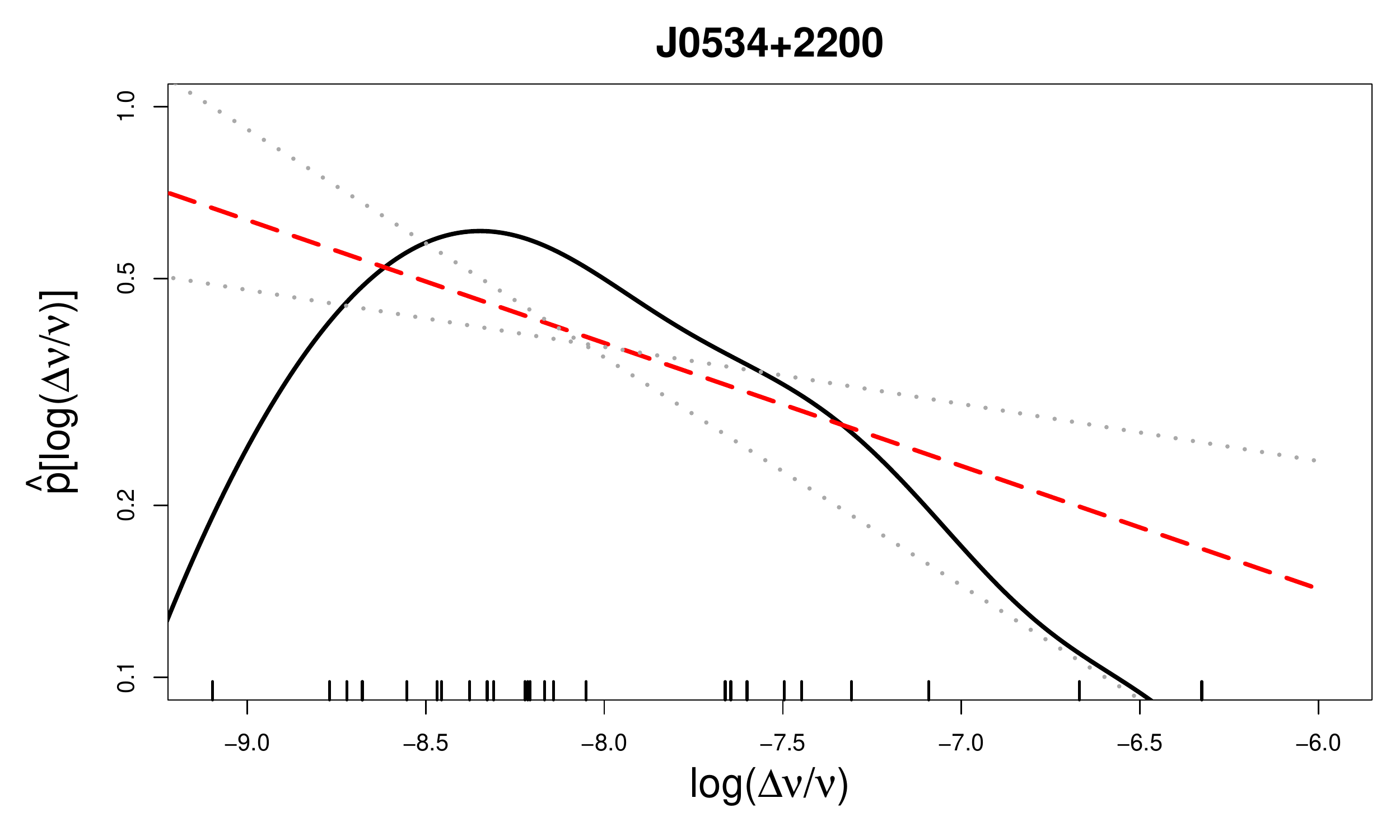} & 
\includegraphics[scale=0.25]{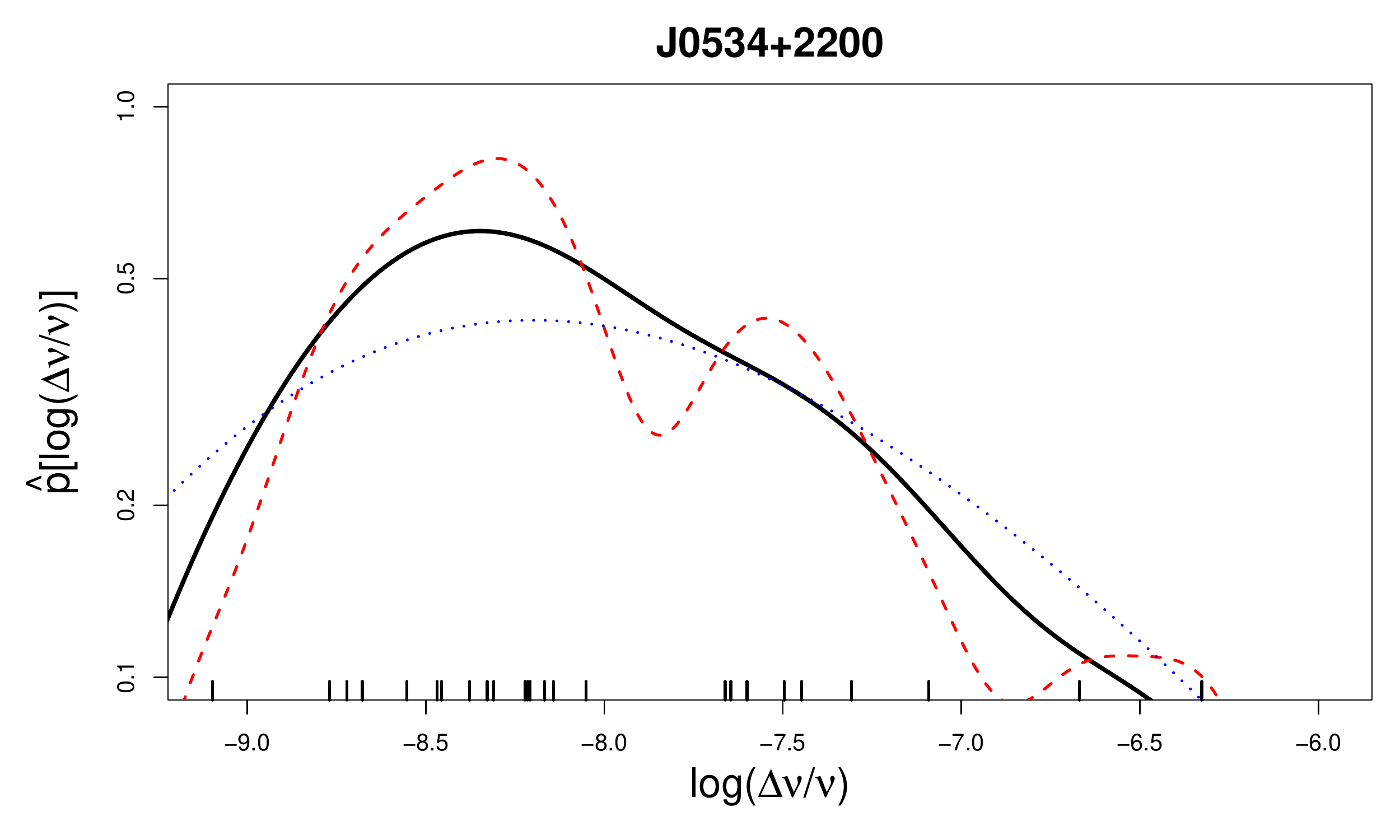} \\
\includegraphics[scale=0.25]{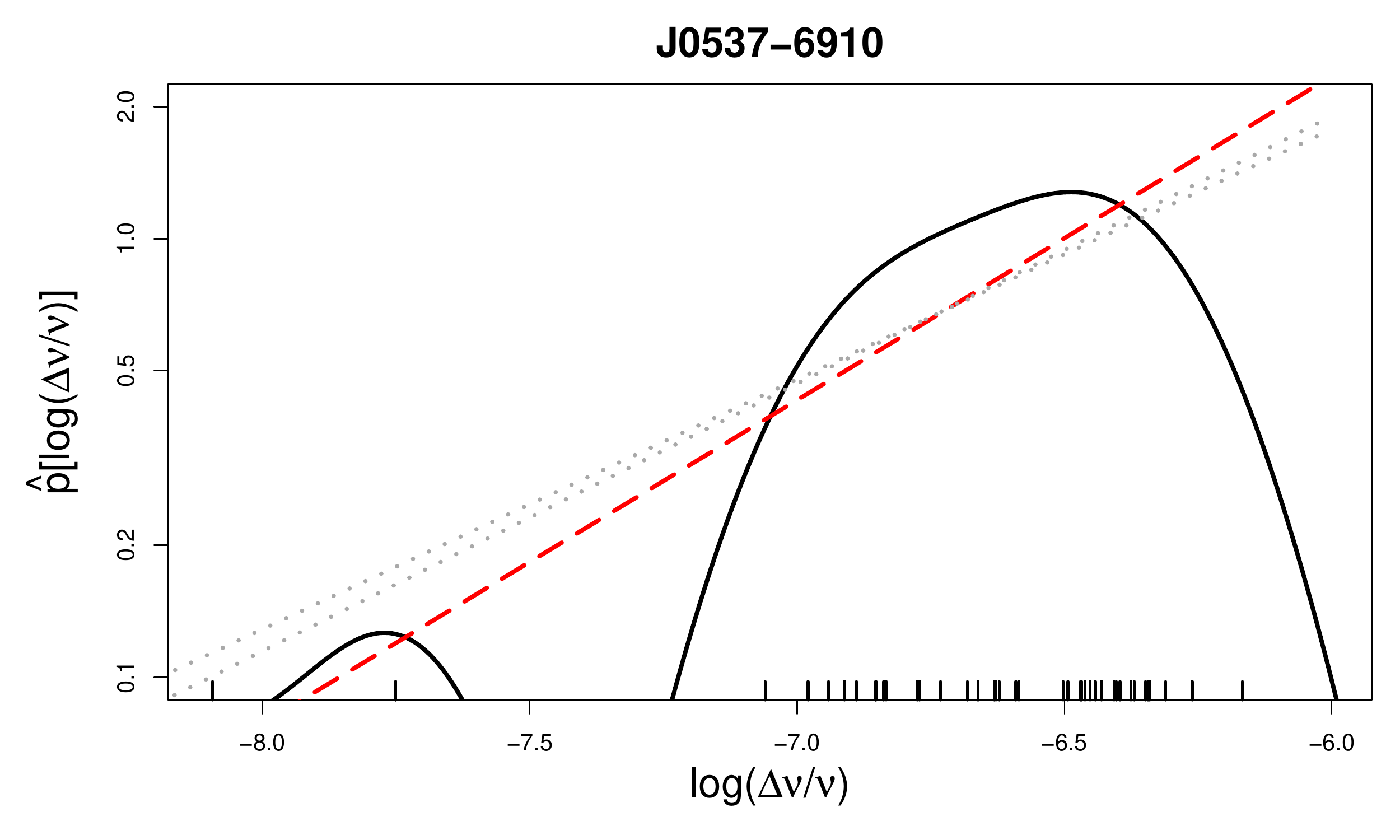} &
\includegraphics[scale=0.25]{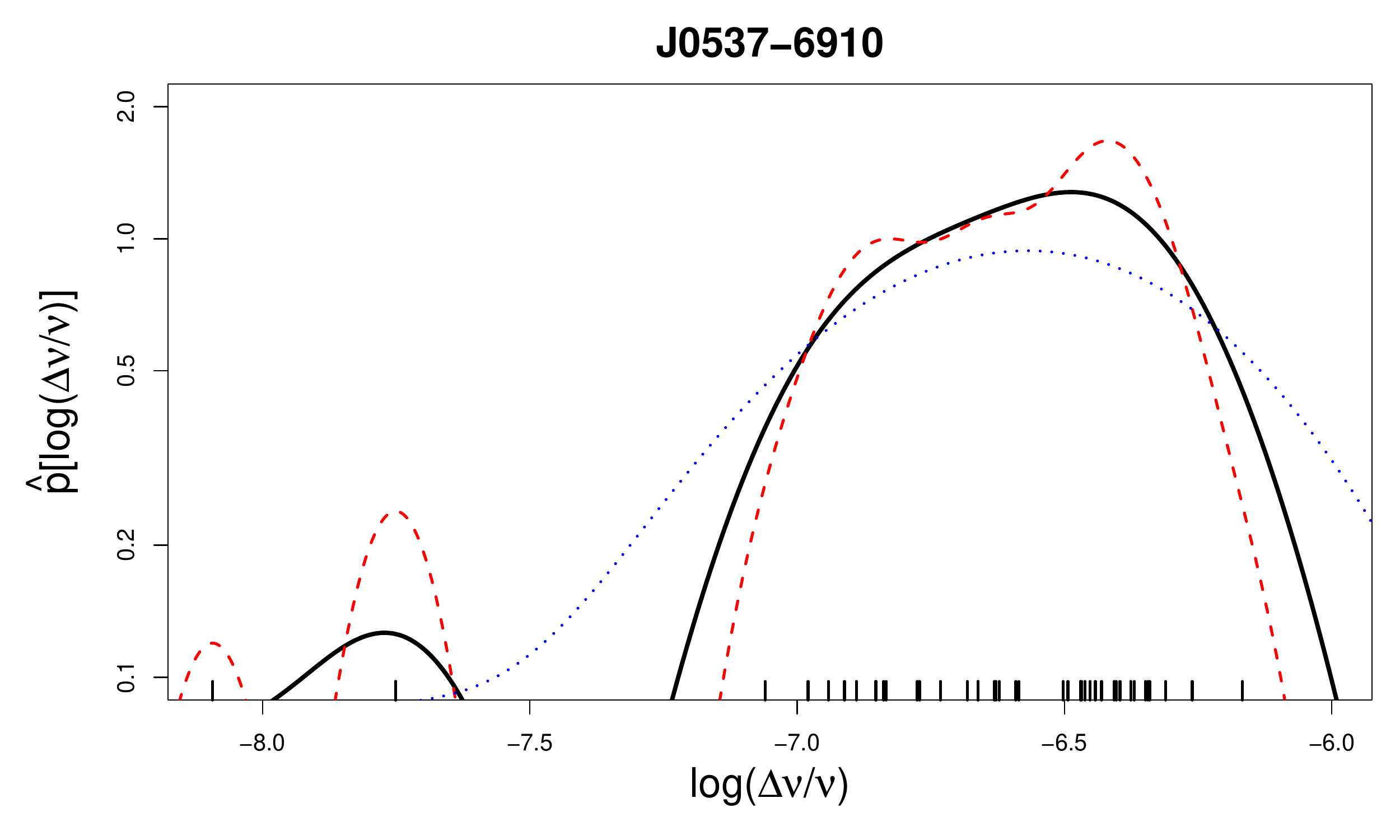} \\
\includegraphics[scale=0.25]{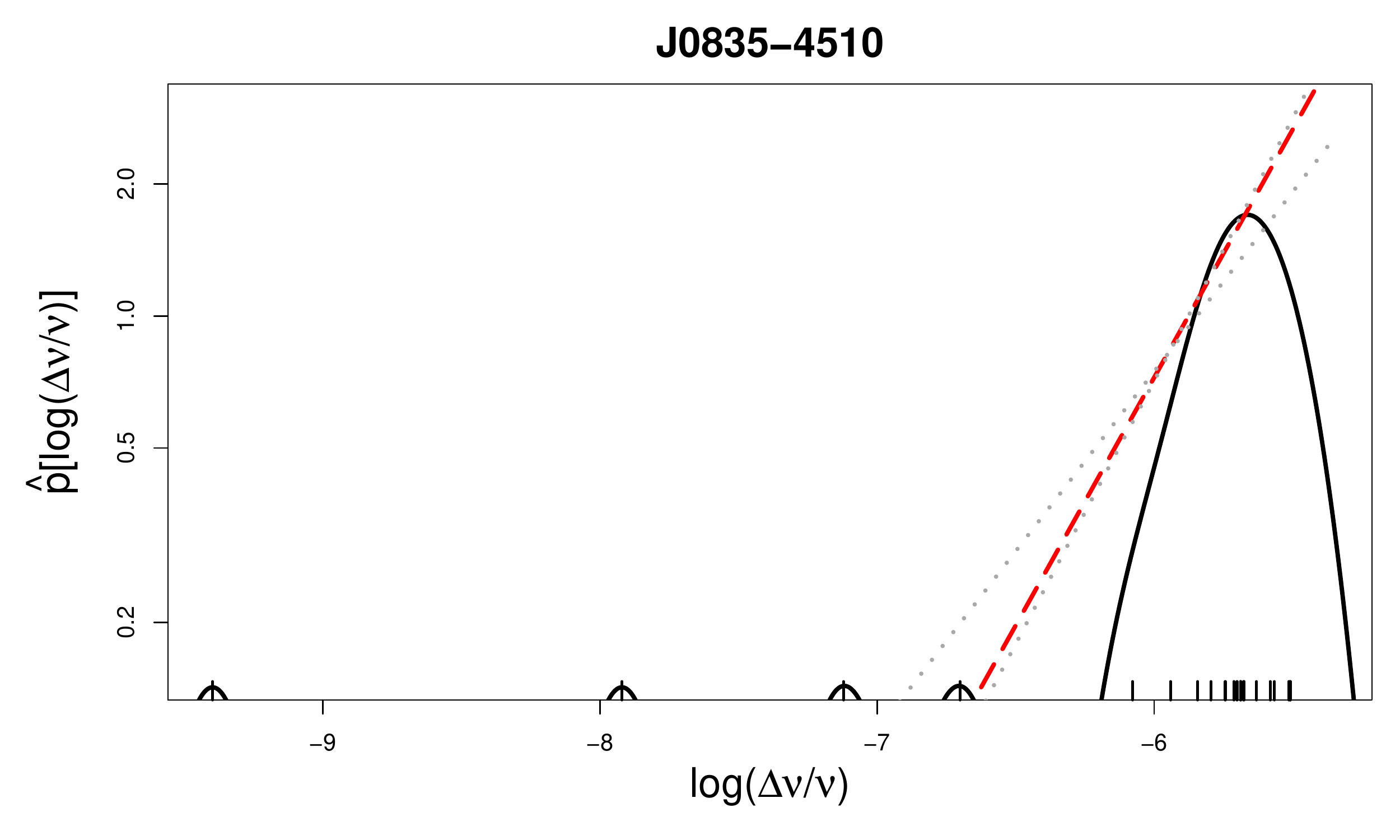} &
\includegraphics[scale=0.25]{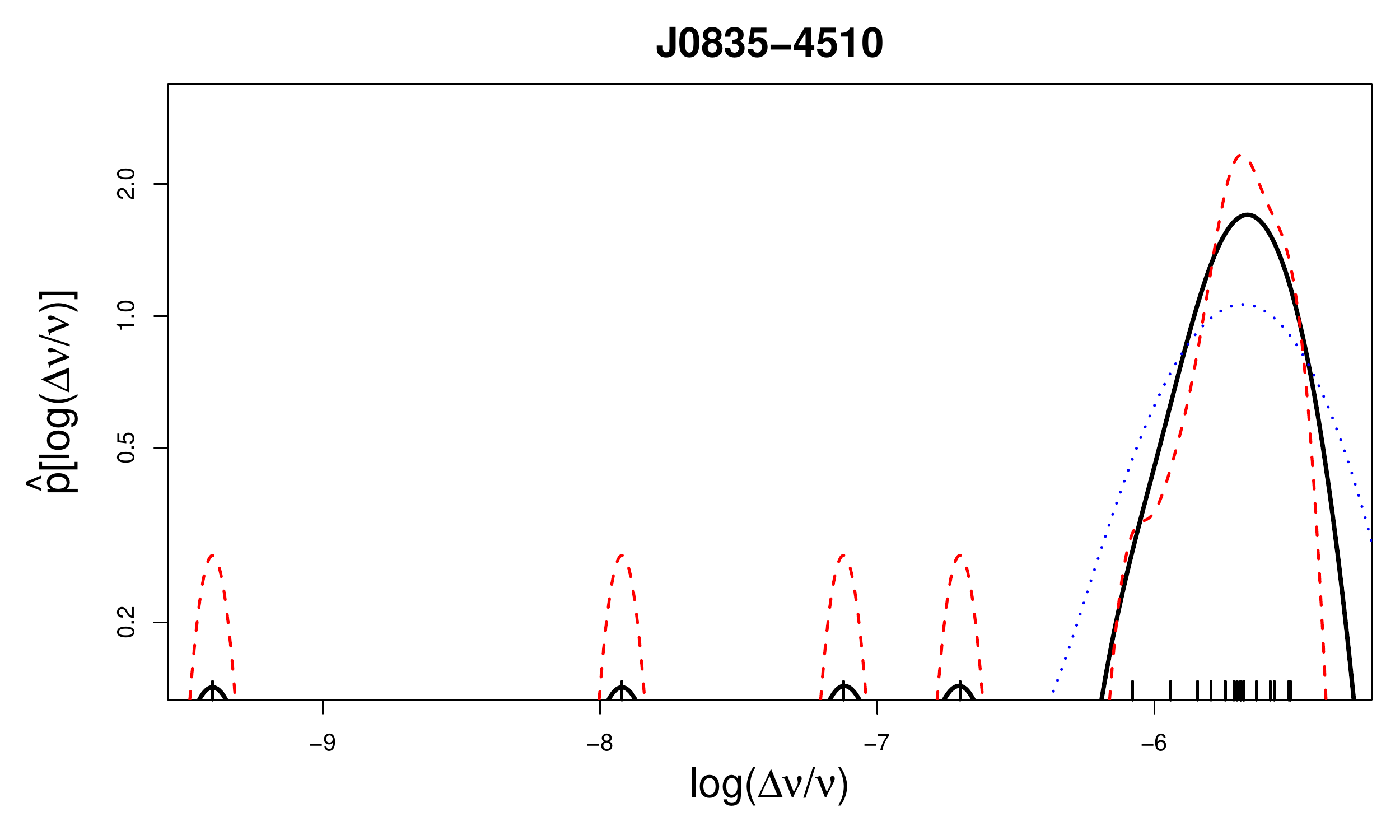} \\
\includegraphics[scale=0.25]{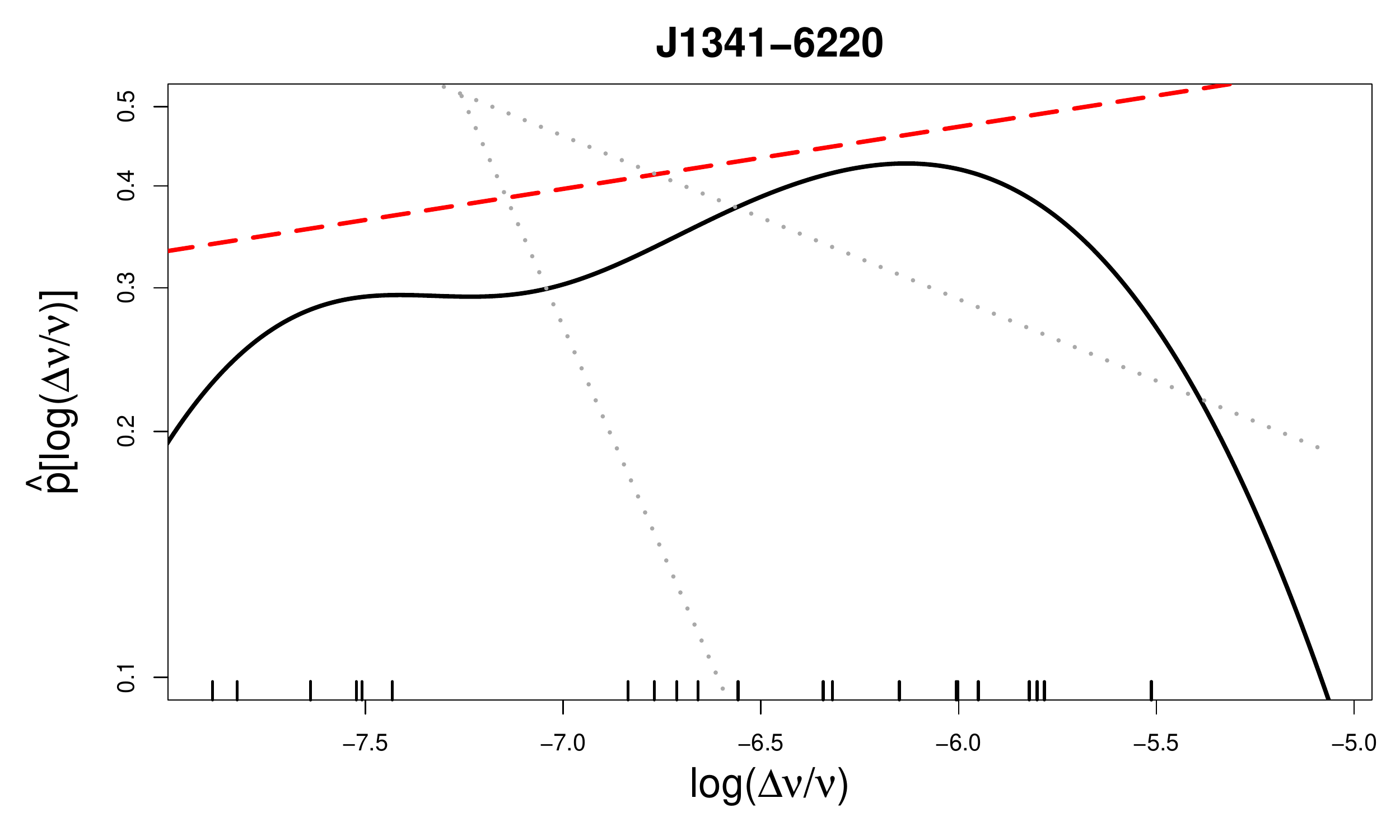} &
\includegraphics[scale=0.25]{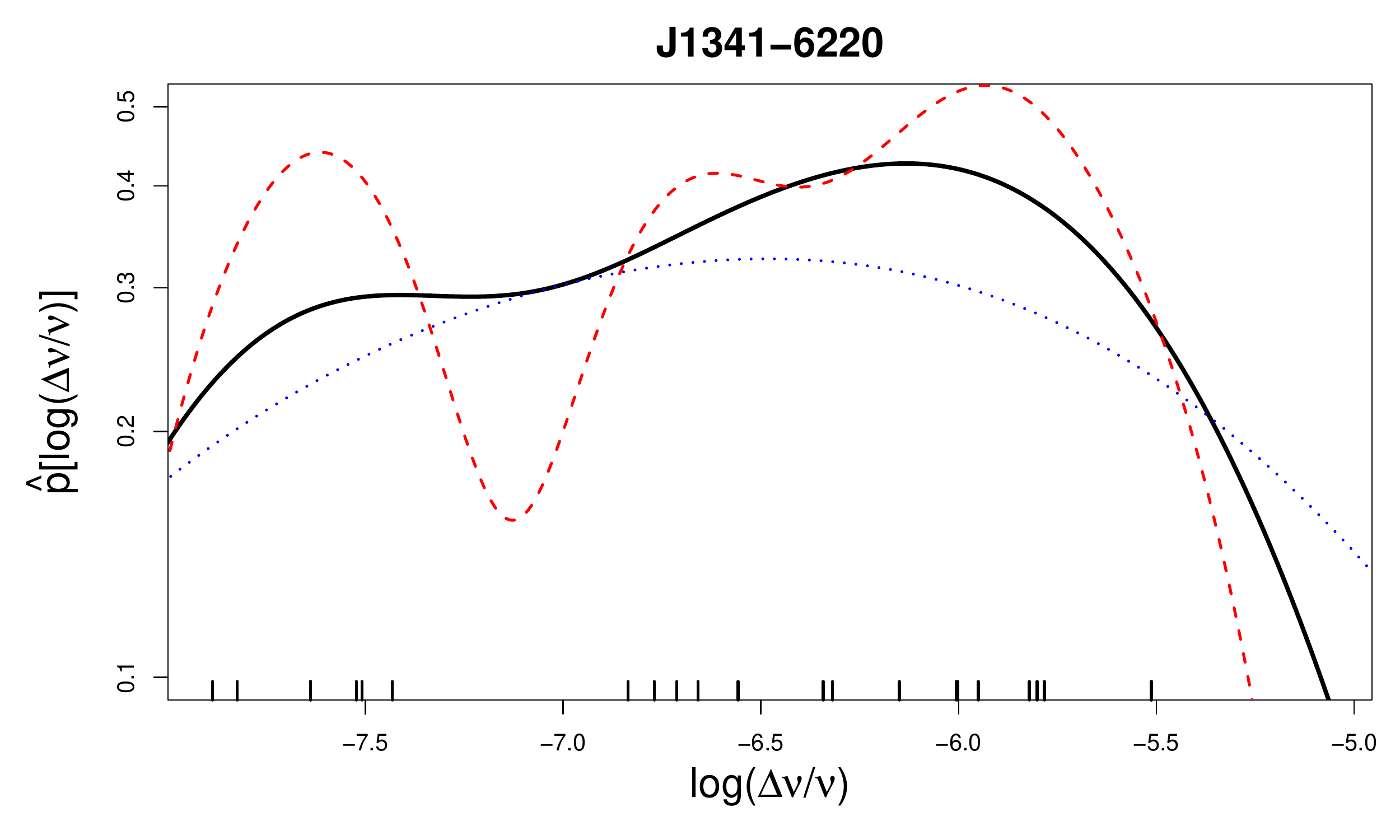} \\
\includegraphics[scale=0.25]{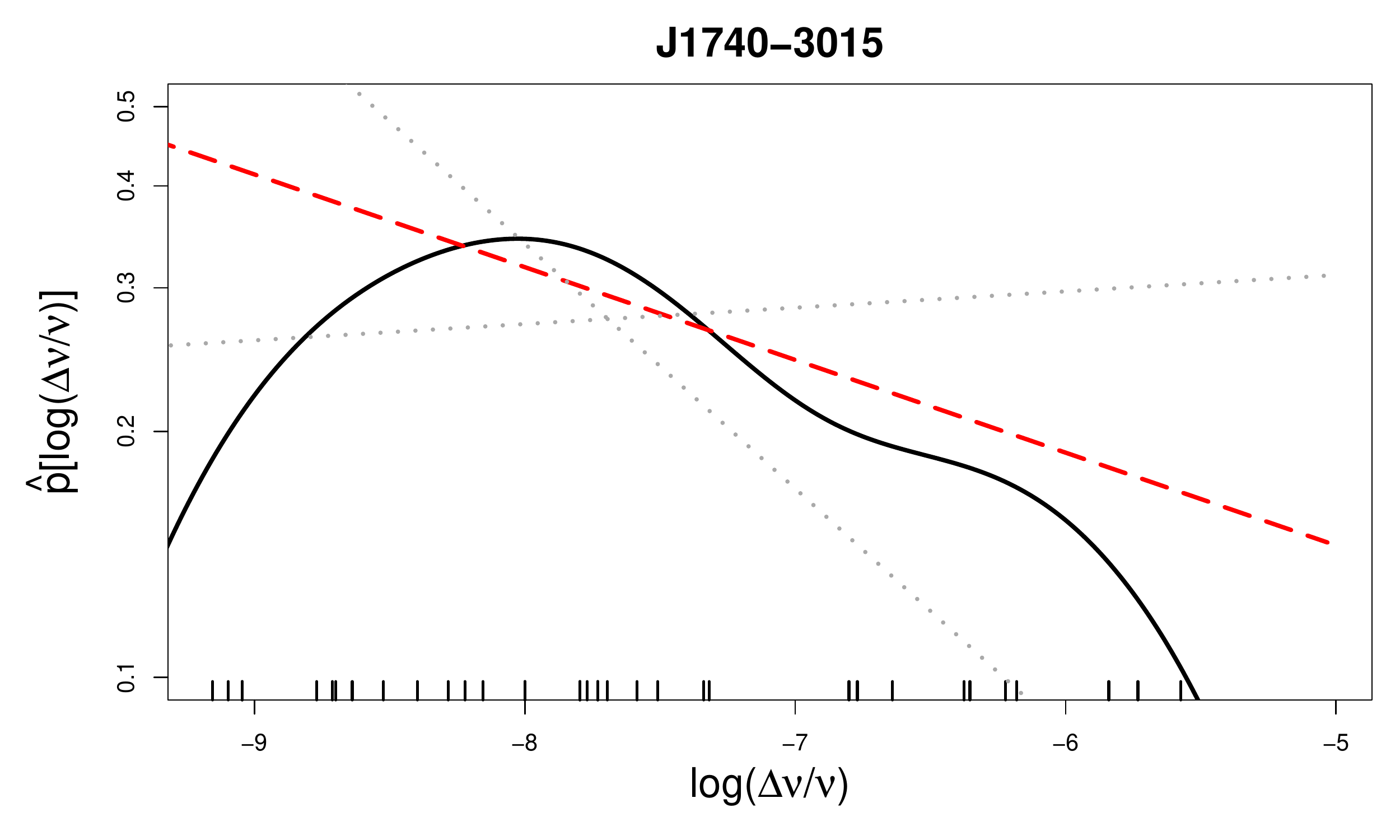} &
\includegraphics[scale=0.25]{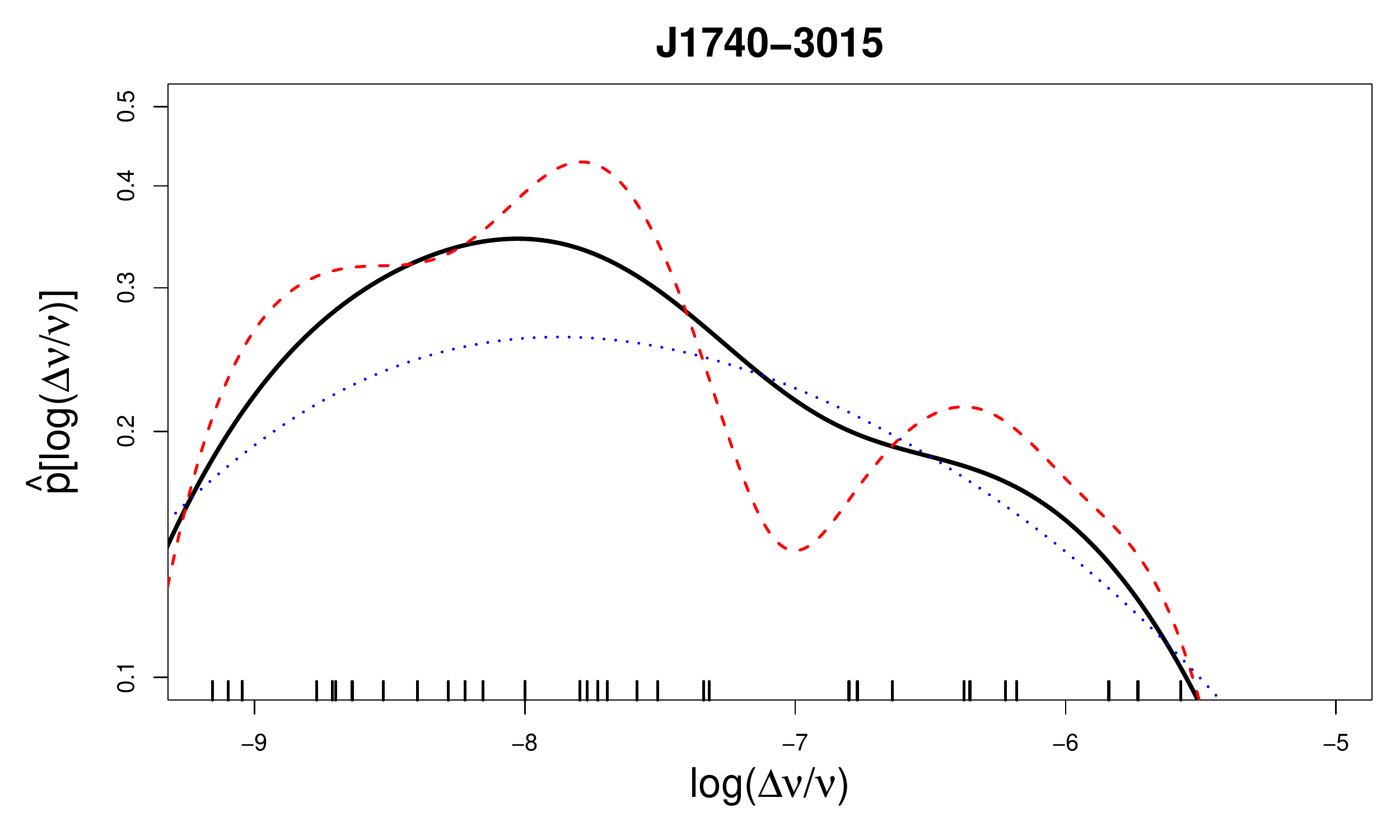}
\end{tabular}
\caption{
Kernel density estimates of the fractional size PDFs ${\hat p}(s)$, 
for the five most active glitchers,
plotted on log-log axes.
The solid black curve is ${\hat p}(s)$ using a normal reference
bandwidth, $h = h_0$.
In the left-hand column, the dashed red lines show the parametric power-law fit 
with $a$ estimated by maximum likelihood.
The grey dotted lines are power-law fits with $a=a_\pm$ from the
K-S analysis in \citet{mel08}.
In the right-hand column, the red dashed curves are ${\hat p}(s)$
with $h = h_0/2$, and the blue dotted curves are ${\hat p}(s)$ with $h=2h_0$.
Tick marks on the horizontal axis indicate the abscissae of the raw data.
}
\label{fig:non4}
\end{figure}

Figure \ref{fig:non4} displays the estimated PDF of the
fractional size logarithm,
$s = \log_{10}(\Delta \nu/\nu)$.
In the left column, the nonparametric estimate ${\hat p}(s)$
from (1) using a normal reference bandwidth, $h = h_0$, is graphed as a solid curve.
Also displayed are parametric fits to the power law distribution
$p(\Delta\nu) = 
(1-a) 
(\Delta\nu_{\rm max}^{1-a} - \Delta\nu_{\rm min}^{1-a})^{-1} 
\Delta\nu^{-a}$ 
The exponent $a$ is estimated by maximum likelihood (red dashed line) 
and by calculating the end-points $a = a_\pm$ (grey dotted lines)
of $a_- \leq a \leq a_+$,
where the K-S probability exceeds 32\% as in Figure \ref{fig:non2};
see \citet{mel08} for details.
There is an art to choosing $\Delta\nu_{\rm min}$ and $\Delta\nu_{\rm max}$,
as discussed in {\S}4.3 in \citet{mel08}.
The smallest glitch observed is likely to be a reasonable estimate
of $\Delta\nu_{\rm min}$, 
because $p(\Delta\nu)$ rises steeply as $\Delta\nu \rightarrow 0$,
but this has not been quantified systematically except by \citet{jan06},
who simulated microglitch detection in a noisy time series and found
$\Delta\nu_{\rm min} = 10^{-10}\nu$ for recent Jodrell Bank observations;
see also \citet{esp14}, where it is claimed that the lower cut-off in
$p(\Delta \nu)$ can be resolved.
Below we set $\Delta\nu_{\rm min}$ and $\Delta\nu_{\rm max}$ 
to the observed minimum and maximum respectively;
the results are insensitive to this choice, as demonstrated previously
\citep{mel08}.
If the glitch size distribution truly follows a power law, then it is bounded
from below and the estimator ${\hat p}(s)$ 
should include a reflection boundary correction as in \S\ref{sec:non2d} and Figure 1.
However, we have neglected to perform this boundary correction, since
we don't know where (if anywhere) the boundary is.
In the right column of Figure \ref{fig:non4}, we show ${\hat p}(s)$ with 
$h = h_0$ as a black solid curve, and also $\hat{p}(s)$ with $h = h_0/2$ (dashed red curve)
and $h = 2h_0$ (dotted blue curve).

Figure \ref{fig:non4} elaborates the classification stemming
from Figure \ref{fig:non3} in two interesting ways.
First,
we find that ${\hat p}(s)$ generally decreases 
with $s$, with a prominent peak at $\approx s_{\rm min}$,
for the Poisson-like duo,
PSR J0534$+$2200 and PSR J1740$-$3015,
whose waiting time PDFs also decrease \footnote{
The drop in the PDF at $s < s_{\rm min}$ is an artifact;
it traces the leftward Gaussian tail of the leftmost kernel.
}.
In contrast,
PSR J0537$-$6910 and PSR J0835$-$4510 are not monotonic,
with ${\hat p}(s_{\rm min})$ less than 20\% of ${\hat p}(s)$ at the peak
and the bulk of the probability confined within $\lesssim 1\,{\rm dex}$.
PSR J1341$-$6220,
previously classified as Poisson-like by \citet{mel08}
and now a quasiperiodic candidate on the basis of Figure \ref{fig:non1},
displays interesting intermediate behavior:
${\hat p}(s)$ does not decrease monotonically
but nor is it narrowly peaked,
staying nearly flat over $\approx 2\,{\rm dex}$.
This may mean one of two important things:
(i) the broad waiting time plateau in Figure \ref{fig:non1} actually extends
to $\Delta t = 0$ and would emerge more clearly given more data,
ultimately putting PSR J1341$-$6220 in the Poisson class;
or (ii) the correspondence between 
power-law size and exponential waiting time statistics is not one-to-one
\citep{mel08}.

Second,
it is apparent by eye that in the two cases where ${\hat p}(s)$ decreases monotonically
(PSR J0534+2200 and PSR J1740$-$3015), ${\hat p}(s)$ is broadly consistent
with a power law.
The dashed line is consistent with the solid curve and is bracketed by
the dotted fits from \citet{mel08}, 
even after adding three events since 2008.
Interestingly, 
the Poisson-like duo have $1.11 \leq a \leq 1.22$
for the best-fitting dashed line,
whereas PSR J1341$-$6220 has $a=0.92$\footnote{Recall from footnote 4 that the
PDF of the logarithm of a power law-distributed variable with power law index 
$a$ is a power law with index $1-a$, and hence has a positive slope for $a<1$.}.
This division is significant physically, because the mean is dominated
by different extremes of the distribution in the two cases,
with 
$\langle \Delta\nu \rangle
 = (a-1)(2-a)^{-1} 
 (\Delta\nu_{\rm max}/\Delta\nu_{\rm min})^{1-a}
 \Delta\nu_{\rm max}$
for $1 < a < 2$
and 
$\langle \Delta\nu \rangle
 = (a-1)(2-a)^{-1}\Delta\nu_{\rm max}$
for $a < 1$
(assuming $\Delta\nu_{\rm min} \ll \Delta\nu_{\rm max}$).

It is sometimes claimed in the literature
that the quasiperiodic glitchers harbor two distinct event populations
\citep{mel08}.
We find no compelling evidence to support this claim in Figure \ref{fig:non4}.
With $h = h_0$, there are weak hints of two peaks
at $s \approx -7.7$ and $s \approx -6.5$
in PSR J0537$-$6910, the larger events being more frequent.
The hypothetical populations are well separated relative to $h$,
but the lesser group contains just four out of 40 events,
which are probably random outliers;
the validation experiments in Figure \ref{fig:non1} routinely
produce a few outliers in any given realization, 
which show up as a small, low-$s$ bump for $N\leq 25$.
Likewise, in PSR J1341$-$6220,
the broad plateau in ${\hat p}(s)$ contains a hint of two bumps
at $s \approx -7.4$ and $s \approx -6.1$,
but again similar noise artifacts are seen in Figure \ref{fig:non1}.
As in Figure \ref{fig:non3}, using $h = h_0/2$ produces multiple spurious
peaks in every object, suggesting this bandwidth is undersmoothing significantly,
while $h = 2h_0$ produces a similarly-shaped estimate with a single peak that is broadly
consistent with the result for $h = h_0$.
More data may alter this picture, but for now the nonparametric case
for bimodality in glitch sizes is weak in the five objects analysed here.

\section{Conclusion
 \label{sec:non4}}
As the number of detected radio pulsar glitches grows,
it is ever more feasible to disaggregate the data
and construct size and waiting time distributions for individual objects.
Previous statistical analyses have assumed theoretically
inspired parametric PDFs (e.g.\ power law, exponential, finite mixture) 
and tested for K-S inconsistency
\citep{mel08,esp11,kon14,onu16}.
In this paper, we estimate the PDFs nonparametrically
using the kernel density estimator,
a powerful tool which converges optimally in many circumstances
and has proved its mettle in many applications,
where data sets are modest in size
\citep{wan95}.

Our analysis yields three main results.
(i) 
It confirms the existence of two classes of glitcher:
Poisson-like objects with monotonically decreasing waiting time and size PDFs,
consistent with exponential and power-law functional forms respectively,
and objects with peaked waiting time and size PDFs,
which trigger quasiperiodically.
(ii) 
It suggests that one object which was previously classified
as Poisson-like,
PSR J1341$-$6220,
shows evidence of hybrid behaviour:
${\hat p}(\Delta t)$ peaks at $\Delta t > 0$,
but the peaks in ${\hat p}(\Delta t)$ and ${\hat p}(s)$ are broader than for 
PSR J0537$-$6910 and PSR J0835$-$4510,
and ${\hat p}(\Delta t)$ as a whole remains marginally consistent with
an exponential distribution in a K-S sense.
(iii)
One sees weak hints of bimodality in 
${\hat p}(\Delta t)$ and ${\hat p}(s)$,
consistent with previous parametric modeling of quasiperiodic glitchers
\citep{mel08}
and a modest excess of large glitches in aggregate statistics
\citep{lyn00}.
However, Monte-Carlo experiments (Figure \ref{fig:non1}) and a sensitivity study
addressing bandwidth selection in \S\ref{sec:non3a} and \S\ref{sec:non3b} present an inconclusive picture.
The putative bimodality is as likely to be a noise artifact as not, 
given the data at hand.

Do the above results shed new light on glitch physics?
This complicated question will be addressed in depth
in a forthcoming theoretical paper.
Here we venture to make one, model-independent point
without favoring any of the microphysical mechanisms
referenced in \S\ref{sec:non1}.
It is thought that glitches are driven by the electromagnetic torque,
which spins down the stellar crust differentially 
with respect to internal components.
Stresses of various kinds
(e.g.\ elastic forces in the context of crust quakes, 
Magnus forces in the context of superfluid vortex dynamics)
build up globally yet inhomogeneously,
until a glitch is triggered,
whereupon they relax locally via
chains (`avalanches') of threshold-activated, stick-slip events
(e.g.\ crust cracking, or vortex unpinning).
The avalanches must be cooperative phenomena mediated by a
knock-on process,
otherwise the central limit theorem predicts
size and waiting time PDFs dramatically narrower than those observed,
given the large number of interacting elements involved
(e.g.\ $\gtrsim 10^{10}$ vortices)
\citep{war12,war13,ful16}.
Avalanche systems of this sort, like sand piles,
tend towards a self-organized critical state and operate in two regimes:
(i) slow driving,
where successive events are triggered in spatially distinct regions
and are mutually independent, i.e.\ Poisson-like,
with scale-invariant size distributions,
and (ii) fast driving,
where successive events involve the whole system
(e.g.\ all the vortices in a superfluid, or all the grains in a sand pile)
and recur quasiperiodically with roughly equal sizes
\citep{jen98,mel08}.
In this sense, therefore, the nonparametric PDF estimates 
in Figures \ref{fig:non3}--\ref{fig:non4}
are broadly consistent with theoretical expectations,
irrespective of the microphysics.
However,
three additional implications flow from the new results in this paper.
First, 
with the hybrid behaviour of PSR J1341$-$6220 defying easy categorization,
it makes sense to look harder for independent ways to classify 
radio pulsars as slowly/rapidly driven in the sense above\footnote{
Preliminary analysis of the next most active glitchers,
PSR J0631$+$1036 ($N=15$) and PSR J1801$-$2304 ($N=13$),
suggests that the former is Poisson-like,
and the latter is quasiperiodic with a broad ${\hat p}(\Delta t)$
like PSR J1341$-$6220.
However, $N$ is too small to interpret the results reliably.
More data are needed and will be forthcoming soon,
as both objects are active,
with
$\langle \Delta t \rangle = 1.1\,{\rm yr}$ and $2.3\,{\rm yr}$ respectively.
}.
Second,
among the quasiperiodic candidates,
${\hat p}(s)$ is considerably narrower
in PSR J0835$-$4510 ($\approx 0.5\,{\rm dex}$)
than in the others ($\approx 2.5\,{\rm dex}$),
and ${\hat p}(\Delta t)$ is relatively broad in
PSR J1341$-$6220
(indeed almost exponential),
raising the interesting possibility that the nexus
between power-law size and exponential waiting time statistics,
characteristic of self-organized criticality,
may sometimes be broken.
Third, 
if multimodality is uncovered in any object in the future,
as more data are collected,
it will argue for the existence of an additional, non-scale-invariant trigger
in the glitch mechanism,
at least in some pulsars.
How this relates (if at all) to the violation of scale-invariance implied by 
the minimum glitch size resolved in PSR J0534+2200
\citep{esp14} is an interesting question for future work.

\section*{Acknowledgements}
The authors record their warm and respectful gratitude to the late 
Professor Peter Gavin Hall at the University of Melbourne 
for his input during the early part of this work, including his 
recommendation to use the KDE technique and his expert guidance 
concerning smoothing and edge correction issues.
This research was supported by the Australian Research Council through
a discovery project grant and the Centre of Excellence for Gravitational 
Wave Discovery (OzGrav; project number CE170100004).

\bibliographystyle{mn2e}
\bibliography{nonparamglitch8}

\end{document}